\documentclass[11pt]{article}%
\usepackage{amsmath}
\usepackage{graphicx}%
\usepackage{amsfonts}%
\usepackage{amssymb}
\newcommand{\eqnb}{\begin{equation}}
\newcommand{\eqne}{\end{equation}}

\newtheorem{The}{Theorem}

\newtheorem{Cor}[The]{Corollary}
\newtheorem{Lem}{Lemma}
\newtheorem{Pro}{Proposition}
\newtheorem{Rem}{Remark}

\begin{document}

\title{\textbf{Doubly Exponential Solutions for Randomized Load Balancing Models with
Markovian Arrival Processes and Phase-Type Service Times}}
\author{Quan-Lin Li${}^{1}$ \hspace{0.1in}John C.S. Lui${}^{2}$\\${}^{1}$ School of Economics and Management Sciences \\Yanshan University, Qinhuangdao 066004, China \\${}^{2}$ Department of Computer Science \& Engineering \\The Chinese University of Hong Kong, Shatin, N.T, Hong Kong}
\maketitle

\begin{abstract}
In this paper, we provide a novel matrix-analytic approach for studying doubly
exponential solutions of randomized load balancing models (also known as
\emph{supermarket models}) with Markovian arrival processes (MAPs) and
phase-type (PH) service times. We describe the supermarket model as a system
of differential vector equations by means of density dependent jump Markov
processes, and obtain a closed-form solution with a doubly exponential
structure to the fixed point of the system of differential vector equations.
Based on this, we show that the fixed point can be decomposed into the product
of two factors inflecting arrival information and service information, and
further find that the doubly exponential solution to the fixed point is not
always unique for more general supermarket models. Furthermore, we analyze the
exponential convergence of the current location of the supermarket model to
its fixed point, and apply the Kurtz Theorem to study density dependent jump
Markov process given in the supermarket model with MAPs and PH service times,
which leads to the Lipschitz condition under which the fraction measure of the
supermarket model weakly converges the system of differential vector
equations. This paper gains a new understanding of how workload probing can
help in load balancing jobs with non-Poisson arrivals and non-exponential
service times.

\vskip                                                      0.5cm


\end{abstract}

\section{Introduction}

Randomized load balancing, where a job is assigned to a server from a small
subset of randomly chosen servers, is very simple to implement, and can
surprisingly deliver better performance (for example reducing collisions,
waiting times, backlogs) in a number of applications, such as data centers,
hash tables, distributed memory machines, path selection in networks, and task
assignment at web servers. One useful model extensively used to study
randomized load balancing schemes is the supermarket model. In the supermarket
model, a key result by Vvedenskaya, Dobrushin and Karpelevich \cite{Vve:1996}
indicated that when each Poisson arriving job is assigned to the shortest one
of $d\geq2$ randomly chosen queues with exponential service times, the
equilibrium queue length can decay doubly exponentially in the limit as the
population size $n\rightarrow\infty$, and the stationary fraction of queues
with at least $k$ customers is $\rho^{\frac{d^{k}-1}{d-1}}$, which indicates a
substantially exponential improvement over the case for $d=1$, where the tail
of stationary queue length in the corresponding M/M/1 queue is $\rho^{k}$. At
the same time, the exponential improvement is also illustrated by another key
work in which Luczak and McDiarmid \cite{Luc:2006} studied the maximum queue
length in the supermarket model with Poisson arrivals and exponential service times.

The distributed load balancing strategies in which individual job decisions
are based on information on a limited number of other processors, have been
studied by analytical methods in Eager, Lazokwska and Zahorjan
\cite{Eag:1986a, Eag:1986b, Eag:1988} and by trace-driven simulations in Zhou
\cite{Zhou:1988}. Based on this, the supermarket models can be developed by
using either queueing theory or Markov processes. Most of recent research
deals with a simple supermarket model with Poisson arrivals and exponential
service times by means of density dependent jump Markov processes. The methods
used in the recent literature are based on determining the behavior of the
supermarket model as its population size grows to infinity, and its behavior
is naturally described as a system of differential equations whose fixed point
leads to a closed-form solution with a doubly exponential structure. Readers
may refer to, such as, Azar, Broder, Karlin and Upfal \cite{Azar:1999},
Vvedenskaya, Dobrushin and Karpelevich \cite{Vve:1996} and Mitzenmacher
\cite{Mit:1996a, Mit:1996b}.

Certain generalizations of the supermarket models have been explored, for
example, in studying simple variations by Mitzenmacher and V\"{o}cking
\cite{Mit:1998}, Mitzenmacher \cite{Mit:1998a, Mit:1999a, Mit:2001},
V\"{o}cking \cite{Voc:1999}, Mitzenmacher, Richa, and Sitaraman
\cite{Mit:2001a} and Vvedenskaya and Suhov \cite{Vve:1997}; in discussing load
information by Mirchandaney, Towsley, and Stankovic \cite{Mir:1989}, Dahlin
\cite{Dah:1999} and Mitzenmacher \cite{Mit:2000, Mit:2001a}; and in
mathematical analysis by Graham \cite{Gra:2000a, Gra:2000b, Gra:2004}, Luczak
and Norris \cite{LucN:2005} and Luczak and McDiarmid \cite{Luc:2006,
Luc:2007}. Using fast Jackson networks, Martin and Suhov \cite{Mar:1999},
Martin \cite{Mar:2001}, Suhov and Vvedenskaya \cite{Suh:2002} studied
supermarket mall models, where each node in a Jackson network is replaced by
$N$ parallel servers, and a job joins the shortest of $d$ randomly chosen
queues at the node to which it is directed. For non-Poisson arrivals or for
non-exponential service times, Li, Lui and Wang \cite{Li:2010a} discussed the
supermarket model with Poisson arrivals and PH service times, and indicated
that the fixed point decreases doubly exponentially, where the stationary
phase-type environment is shown to be a crucial factor. Bramson, Lu and
Prabhakar \cite{Bra:2010} provided a modularized program based on ansatz for
treating the supermarket model with Poisson arrivals and general service
times, and Li \cite{Li:2010b} further discussed this supermarket model by
means of a system of integral-differential equations, and illustrated that the
fixed point decreases doubly exponentially and that the heavy-tailed service
times do not change the doubly exponential solution to the fixed point.

For the PH distribution, readers may refer to Neuts \cite{Neu:1981, Neu:1989}
and Li \cite{Li:2010}. The MAP is a useful mathematical model, for example,
for describing bursty traffic, self similarity and long-range dependence in
modern computer networks, e.g., see Adler, Feldman and Taqqu \cite{Adl:1998}.
For detail information of the MAP, readers may refer to Chapter 5 in Neuts
\cite{Neu:1989}, Lucantoni \cite{Luc:1991}, Chapter 1 in Li \cite{Li:2010},
and three excellent overviews by Neuts \cite{Neu:1995}, Chakravarthy
\cite{Cha:2000} and Cordeiro and Kharoufeh \cite{Cor:2009}. In computer
networks, Andersen and Nielsen \cite{And:1998} applied the MAP to describe
long-range dependence, and Yoshihara, Kasahara and Takahashi \cite{Yos:2001}
analyzed self-similar traffic by means of a Markov-modulated Poisson process.

It is interesting to answer whether or how non-Poisson arrivals or
non-exponential service times can disrupt doubly exponential solutions to the
fixed points in supermarket models. To that end, this paper studies a
supermarket model with MAPs and PH service times, and shows that there still
exists a doubly exponential solution to the fixed point. The main
contributions of the paper are threefold. The first one is to provide a novel
matrix-analytic approach to study the supermarket model with MAPs and PH
service times. Based on density dependent jump Markov processes, the
supermarket model is described as a system of differential vector equations
whose fixed point has a closed-form solution with a doubly exponential
structure. The second one is to obtain a crucial result that the fixed point
can be decomposed into the product of two factors inflecting arrival
information and service information, which indicates that the doubly
exponential solution to the fixed point can exist extensively, but it is not
always unique for more general supermarket models. The third one is to analyze
exponential convergence of the current location of the supermarket model to
its fixed point. Not only does the exponential convergence indicate the
existence of the fixed point, but it also shows that such a convergent process
is very fast. To study the limit behavior of the supermarket model as its
population size goes to infinity, this paper applies the Kurtz Theorem to
study density dependent jump Markov process given in the supermarket model
with MAPs and PH service times, which leads to the Lipschitz condition under
which the fraction measure of the supermarket model weakly converges the
system of differential vector equations.

The remainder of this paper is organized as follows. In Section 2, we first
describe a supermarket model with MAPs and PH service times. Then the
supermarket model is described as a systems of differential vector equations
in terms of density dependent jump Markov processes. In Section 3, we first
introduce a fixed point of the system of differential vector equations, and
set up a system of nonlinear equations satisfied by the fixed point. Then we
provide a closed-form solution with a doubly exponential structure to the
fixed point, and show that the fixed point can be decomposed into the product
of two factors inflecting arrival information and service information. In
Section 4, we provide an important observation in which the doubly exponential
solution to the fixed point is not always unique for more general supermarket
models. In Section 5, we study exponential convergence of the current location
of the supermarket model to its fixed point. In Section 6, we apply the Kurtz
Theorem to study density dependent jump Markov process given in the
supermarket model with MAPs and PH service times, which leads to the Lipschitz
condition under which the fraction measure of the supermarket model weakly
converges the system of differential vector equations. Some concluding remarks
are given in Section 7.

\section{Supermarket Model Description}

In this section, we first provide a supermarket model with MAPs and PH service
times. Then the supermarket model is described as a system of differential
vector equations based on density dependent jump Markov processes.

We first introduce some notation as follows. Let $A\otimes B$ be the
\emph{Kronecker product} of two matrices $A=(a_{i,j})$ and $B=(b_{i,j})$, that
is, $A\otimes B=(a_{i,j}B)$; $A\oplus B$ the \emph{Kronecker sum} of $A$ and
$B$, that is, $A\oplus B=A\otimes I+I\otimes B$. We denote by $A\odot B$ the
\emph{Hadamard Product} of $A$ and $B$ as follows:
\[
A\odot B=\left(  a_{i,j}b_{i,j}\right)  .
\]
Specifically, for $k\geq2$, we have%
\[
A^{\odot k}=\underset{k\text{ matrix }A}{\underbrace{A\odot A\odot\cdots\odot
A}}.
\]
For a vector $a=\left(  a_{1},a_{2},\ldots,a_{m}\right)  $, we write%
\[
a^{\odot\frac{1}{d}}=\left(  a_{1}^{\frac{1}{d}},a_{2}^{\frac{1}{d}}%
,\ldots,a_{m}^{\frac{1}{d}}\right)  .
\]

Now, we describe the supermarket model, which is abstracted as a multi-server
multi-queue queueing system. Customers arrive at a queueing system of $n>1$
servers as a MAP with an irreducible matrix descriptor $\left(  nC,nD\right)
$ of size $m_{A}$. Let $\gamma$ be the stationary probability vector of the
irreducible Markov chain $C+D$. Then the stationary arrival rate of the MAP is
given by $n\lambda=n\gamma De$, where $e$ is a column vector of ones with a
suitable size. The service time of each customer is of phase type with an
irreducible representation $\left(  \alpha,T\right)  $ of order $m_{B}$, where
the row vector $\alpha$ is a probability vector whose $j$th entry is the
probability that a service begins in phase $j$ for $1\leq j\leq m_{B}$; $T$ is
an $m_{B}\times m_{B}$ matrix whose $\left(  i,j\right)  ^{th}$ entry is
denoted by $t_{i,j}$ with $t_{i,i}<0$ for $1\leq i\leq m_{B}$, and
$t_{i,j}\geq0$ for $1\leq i,j\leq m_{B}$ and $i\neq j$. Let $T^{0}%
=-Te\gvertneqq0$. When a PH service time is in phase $i$, the transition rate
from phase $i$ to phase $j$ is $t_{i,j}$, the service completion rate is
$t_{i}^{0}$, and the output rate from phase $i$ is $\mu_{i}=-t_{i,i}$. At the
same time, the expected service time is given by $1/\mu=-\alpha T^{-1}e$. Each
arriving customer chooses $d\geq1$ servers independently and uniformly at
random from the $n$ servers, and waits for service at the server which
currently contains the fewest number of customers. If there is a tie, servers
with the fewest number of customers will be chosen randomly. All customers in
every server will be served in the first-come-first service (FCFS) manner. We
assume that all the random variables defined above are independent, and that
this system is operating in the region $\rho=\lambda/\mu<1$. Please see Figure
1 for an illustration of such a supermarket model.

\begin{figure}[ptbh]
\centering   \includegraphics[width=10cm]{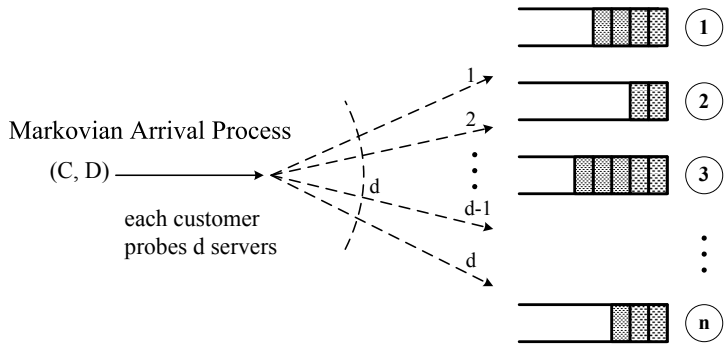}  \caption{The supermarket
model wherein each customer can probe $d$ servers}%
\label{figure: model}%
\end{figure}

The following lemma, which is stated without proof, provides an intuitively
sufficient condition under which the supermarket model is stable. Note that
this proof can be given by a simple comparison argument with the queueing
system in which each customer queues at a random server (i.e., where $d=1$).
When $d=1$, each server acts like a MAP/PH/1 queue which is stable if
$\rho=\lambda/\mu<1$, see chapter 5 in Neuts \cite{Neu:1989}. The comparison
argument is similar to those in Winston \cite{Win:1977} and Weber
\cite{Web:1978}, thus we can obtain two useful results: (1) the shortest queue
is optimal due to the assumptions on MAPs and PH service times; and (2) the
size of the longest queue in the supermarket model is stochastically dominated
by the size of the longest queue in a set of $n$ independent MAP/PH/1 queues.

\begin{Lem}
The supermarket model with MAPs and PH service times is stable if
$\rho=\lambda/\mu<1.$
\end{Lem}

We define $n_{k}^{\left(  i,j\right)  }\left(  t\right)  $ as the number of
queues with at least $k$ customers who include the customer in service, the
MAP in phase $i$ and the PH service time in phase $j$\ at time $t\geq0$.
Clearly, $0\leq n_{k}^{\left(  i,j\right)  }\left(  t\right)  \leq n$ for
$1\leq i\leq m_{A}$, $1\leq j\leq m_{B}$ and $k\geq0$. Let%
\[
x_{n}^{\left(  i\right)  }\left(  0,t\right)  =\frac{n_{0}^{\left(  i\right)
}\left(  t\right)  }{n}%
\]
and for $k\geq1$%
\[
x_{n}^{\left(  i,j\right)  }\left(  k,t\right)  =\frac{n_{k}^{\left(
i,j\right)  }\left(  t\right)  }{n},
\]
which is the fraction of queues with at least $k$ customers, the MAP in phase
$i$ and the PH service time in phase $j$\ at time $t\geq0$. Using the
lexicographic order we write%
\[
X_{n}\left(  0,t\right)  =\left(  x_{n}^{\left(  1\right)  }\left(
0,t\right)  ,x_{n}^{\left(  2\right)  }\left(  0,t\right)  ,\ldots
,x_{n}^{\left(  m_{A}\right)  }\left(  0,t\right)  \right)
\]
and for $k\geq1$%
\begin{align*}
X_{n}\left(  k,t\right)  =  &  (x_{n}^{\left(  1,1\right)  }\left(
k,t\right)  ,x_{n}^{\left(  1,2\right)  }\left(  k,t\right)  ,\ldots
,x_{n}^{\left(  1,m_{B}\right)  }\left(  k,t\right)  ;\ldots;\\
&  x_{n}^{\left(  m_{A},1\right)  }\left(  k,t\right)  ,x_{n}^{\left(
m_{A},2\right)  }\left(  k,t\right)  ,\ldots,x_{n}^{\left(  m_{A}%
,m_{B}\right)  }\left(  k,t\right)  ),
\end{align*}%
\[
X_{n}\left(  t\right)  =\left(  X_{n}\left(  0,t\right)  ,X_{n}\left(
1,t\right)  ,X_{n}\left(  2,t\right)  ,\ldots\right)  .
\]
The state of the supermarket model may be described by the vector
$X_{n}\left(  t\right)  $ for $t\geq0$. Since the arrival process to the
queueing system is a MAP and the service time of each customer is of phase
type, the stochastic process $\left\{  X_{n}\left(  t\right)  ,t\geq0\right\}
$ describing the state of the supermarket model is a Markov process whose
state space is given by%
\begin{align*}
\Omega_{n}=  &  \{\left(  g_{n}^{\left(  0\right)  },g_{n}^{\left(  1\right)
},g_{n}^{\left(  2\right)  }\ldots\right)  :g_{n}^{\left(  0\right)  }\text{
is a probability vector, }g_{n}^{\left(  k-1\right)  }\geq g_{n}^{\left(
k\right)  }\geq0\\
&  \text{ for }k\geq2\text{,}\text{ and }ng_{n}^{\left(  l\right)  }\text{ is
a vector of nonnegative integers for }l\geq0\}.
\end{align*}
Let%
\[
s_{0}^{\left(  i\right)  }\left(  n,t\right)  =E\left[  x_{0}^{\left(
i\right)  }\left(  n,t\right)  \right]
\]
and $k\geq1$%
\[
s_{k}^{\left(  i,j\right)  }\left(  n,t\right)  =E\left[  x_{k}^{\left(
i,j\right)  }\left(  n,t\right)  \right]  .
\]
Using the lexicographic order we write%
\[
S_{0}\left(  n,t\right)  =\left(  s_{0}^{\left(  1\right)  }\left(
n,t\right)  ,s_{0}^{\left(  2\right)  }\left(  n,t\right)  ,\ldots
,s_{0}^{\left(  m_{A}\right)  }\left(  n,t\right)  \right)
\]
and for $k\geq1$%
\begin{align*}
S_{k}\left(  n,t\right)  =  &  (s_{k}^{\left(  1,1\right)  }\left(
n,t\right)  ,s_{k}^{\left(  1,2\right)  }\left(  n,t\right)  ,\ldots
,s_{k}^{\left(  1,m_{B}\right)  }\left(  n,t\right)  ;\ldots;\\
&  s_{k}^{\left(  m_{A},1\right)  }\left(  n,t\right)  ,s_{k}^{\left(
m_{A},2\right)  }\left(  n,t\right)  ,\ldots,s_{k}^{\left(  m_{A}%
,m_{B}\right)  }\left(  n,t\right)  ),
\end{align*}%
\[
S\left(  n,t\right)  =\left(  S_{0}\left(  n,t\right)  ,S_{1}\left(
n,t\right)  ,S_{2}\left(  n,t\right)  ,\ldots\right)  .
\]
As shown in Martin and Suhov \cite{Mar:1999} and Luczak and McDiarmid
\cite{Luc:2006}, the Markov process $\left\{  X_{n}\left(  t\right)
,t\geq0\right\}  $ is asymptotically deterministic as $n\rightarrow\infty$.
Thus the limits $\lim_{n\rightarrow\infty}E\left[  x_{0}^{\left(  i\right)
}\left(  n,t\right)  \right]  $ and $\lim_{n\rightarrow\infty}E\left[
x_{k}^{\left(  i,j\right)  }\left(  n,t\right)  \right]  $ always exist by
means of the law of large numbers. Based on this, we write%
\[
S_{0}\left(  t\right)  =\lim_{n\rightarrow\infty}S_{0}\left(  n,t\right)  ,
\]
for $k\geq1$%
\[
S_{k}\left(  t\right)  =\lim_{n\rightarrow\infty}S_{k}\left(  n,t\right)  ,
\]
and%
\[
S\left(  t\right)  =\left(  S_{0}\left(  t\right)  ,S_{1}\left(  t\right)
,S_{2}\left(  t\right)  ,\ldots\right)  .
\]
Note that $S_{0}\left(  t\right)  $ and $S_{k}\left(  t\right)  $ are two row
vectors of order $m_{A}$ and $m_{A}m_{B}$, respectively. Let $X\left(
t\right)  =\lim_{n\rightarrow\infty}X_{n}\left(  t\right)  $. Then it is easy
to see from the MAPs and the PH service times that $\left\{  X\left(
t\right)  ,t\geq0\right\}  $ is also a Markov process whose state space is
given by%
\[
\Omega=\left\{  \left(  g^{\left(  0\right)  },g^{\left(  1\right)
},g^{\left(  2\right)  },\ldots\right)  :g^{\left(  0\right)  }\text{ is a
probability vector},g^{\left(  k-1\right)  }\geq g^{\left(  k\right)  }%
\geq0\right\}  .
\]
If the initial distribution of the Markov process $\left\{  X_{n}\left(
t\right)  ,t\geq0\right\}  $ approaches the Dirac delta-measure concentrated
at a point $g\in$ $\Omega$, then the limit $X\left(  t\right)  =\lim
_{n\rightarrow\infty}X_{n}\left(  t\right)  $ is concentrated on the
trajectory $S_{g}=\left\{  S\left(  t\right)  :t\geq0\right\}  $. This
indicates a law of large numbers for the time evolution of the fraction of
queues of different lengths. Furthermore, the Markov process $\left\{
X_{n}\left(  t\right)  ,t\geq0\right\}  $ converges weakly to the fraction
vector $S\left(  t\right)  =\left(  S_{0}\left(  t\right)  ,S_{1}\left(
t\right)  ,S_{2}\left(  t\right)  ,\ldots\right)  $ as $n\rightarrow\infty$,
or for a sufficiently small $\varepsilon>0$,%
\[
\lim_{n\rightarrow\infty}P\left\{  ||X_{n}\left(  t\right)  -S\left(
t\right)  ||\geq\varepsilon\right\}  =0,
\]
where $||a||$ is the $L_{\infty}$-norm of vector $a$.

The following proposition shows that the sequence $\left\{  S_{k}\left(
t\right)  ,k\geq0\right\}  $ is monotone decreasing, while its proof is easy
by means of the definition of $S\left(  t\right)  $.

\begin{Pro}
For $1\leq k<l$%
\[
S_{l}\left(  t\right)  <S_{k}\left(  t\right)
\]
and%
\[
S_{l}\left(  t\right)  e<S_{k}\left(  t\right)  e<S_{0}\left(  t\right)  e=1.
\]
\end{Pro}

In what follows we set up a system of differential vector equations satisfied
by the fraction vector $S\left(  t\right)  $ by means of density dependent
jump Markov processes.

We first provide an example to indicate how to derive the differential vector
equations. Consider the supermarket model with $n$ servers, and determine the
expected change in the number of queues with at least $k$ customers over a
small time period of length d$t$. The probability vector that an arriving
customer joins a queue of size $k-1$ during this time period is given by%
\[
\left[  S_{k-1}^{\odot d}\left(  n,t\right)  \left(  D\otimes I\right)
+S_{k}^{\odot d}\left(  n,t\right)  \left(  C\otimes I\right)  \right]  \cdot
n\text{d}t,
\]
since each arriving customer chooses $d\geq1$ servers independently and
uniformly at random from the $n$ servers, and waits for service at the server
which currently contains the fewest number of customers. Similarly, the
probability vector that a customer leaves a server queued by $k$ customers
during this time period is given by%
\[
\left[  S_{k}\left(  n,t\right)  \left(  I\otimes T\right)  +S_{k+1}\left(
n,t\right)  \left(  I\otimes T^{0}\alpha\right)  \right]  \cdot n\text{d}t.
\]
Therefore, we can obtain%
\begin{align*}
\text{d}E\left[  n_{k}\left(  n,t\right)  \right]  = &  \left[  S_{k-1}^{\odot
d}\left(  n,t\right)  \left(  D\otimes I\right)  +S_{k}^{\odot d}\left(
n,t\right)  \left(  C\otimes I\right)  \right]  \cdot n\text{d}t\\
&  +\left[  S_{k}\left(  n,t\right)  \left(  I\otimes T\right)  +nS_{k+1}%
\left(  n,t\right)  \left(  I\otimes T^{0}\alpha\right)  \right]  \cdot
n\text{d}t,
\end{align*}
which leads to%
\begin{align}
\frac{\text{d}S_{k}\left(  n,t\right)  }{\text{d}t}= &  S_{k-1}^{\odot
d}\left(  n,t\right)  \left(  D\otimes I\right)  +S_{k}^{\odot d}\left(
n,t\right)  \left(  C\otimes I\right)  \nonumber\\
&  +S_{k}\left(  n,t\right)  \left(  I\otimes T\right)  +S_{k+1}\left(
n,t\right)  \left(  I\otimes T^{0}\alpha\right)  .\label{Equ1}%
\end{align}
Using a similar analysis for Equation (\ref{Equ1}), we can obtain a system of
differential vector equations for the fraction vector $S\left(  n,t\right)
=\left(  S_{0}\left(  n,t\right)  ,S_{1}\left(  n,t\right)  ,S_{2}\left(
n,t\right)  ,\ldots\right)  $ as follows:%
\begin{equation}
S_{0}\left(  n,t\right)  e=1,\label{E1}%
\end{equation}%
\begin{equation}
\frac{\mathtt{d}}{\text{d}t}S_{0}\left(  n,t\right)  =S_{0}^{\odot d}\left(
n,t\right)  C+S_{1}\left(  n,t\right)  \left(  I\otimes T^{0}\right)
,\label{E2}%
\end{equation}%
\begin{align}
\frac{\mathtt{d}}{\text{d}t}S_{1}\left(  n,t\right)  = &  S_{0}^{\odot
d}\left(  n,t\right)  \left(  D\otimes\alpha\right)  +S_{1}^{\odot d}\left(
n,t\right)  \left(  C\otimes I\right)  \nonumber\\
&  +S_{1}\left(  n,t\right)  \left(  I\otimes T\right)  +S_{2}\left(
n,t\right)  \left(  I\otimes T^{0}\alpha\right)  ,\label{E3}%
\end{align}
and for $k\geq2$%
\begin{align}
\frac{\text{d}S_{k}\left(  n,t\right)  }{\text{d}t}= &  S_{k-1}^{\odot
d}\left(  n,t\right)  \left(  D\otimes I\right)  +S_{k}^{\odot d}\left(
n,t\right)  \left(  C\otimes I\right)  \nonumber\\
&  +S_{k}\left(  n,t\right)  \left(  I\otimes T\right)  +S_{k+1}\left(
n,t\right)  \left(  I\otimes T^{0}\alpha\right)  .\label{E4}%
\end{align}
Noting that the limit $\lim_{n\rightarrow\infty}S_{k}\left(  n,t\right)  $
exists for $k\geq0$ and taking $n\rightarrow\infty$ in the both sides of the
system of differential vector equations (\ref{E1}) to (\ref{E4}), we can
easily obtain a system of differential vector equations for the fraction
vector $S\left(  t\right)  =\left(  S_{0}\left(  t\right)  ,S_{1}\left(
t\right)  ,S_{2}\left(  t\right)  ,\ldots\right)  $ as follows:%
\begin{equation}
S_{0}\left(  t\right)  e=1,\label{Equ3}%
\end{equation}%
\begin{equation}
\frac{\mathtt{d}}{\text{d}t}S_{0}\left(  t\right)  =S_{0}^{\odot d}\left(
t\right)  C+S_{1}\left(  t\right)  \left(  I\otimes T^{0}\right)
,\label{Equ4}%
\end{equation}%
\begin{align}
\frac{\mathtt{d}}{\text{d}t}S_{1}\left(  t\right)  = &  S_{0}^{\odot d}\left(
t\right)  \left(  D\otimes\alpha\right)  +S_{1}^{\odot d}\left(  t\right)
\left(  C\otimes I\right)  \nonumber\\
&  +S_{1}\left(  t\right)  \left(  I\otimes T\right)  +S_{2}\left(  t\right)
\left(  I\otimes T^{0}\alpha\right)  ,\label{Equ5}%
\end{align}
and for $k\geq2$,%
\begin{align}
\frac{\text{d}S_{k}\left(  t\right)  }{\text{d}t}= &  S_{k-1}^{\odot d}\left(
t\right)  \left(  D\otimes I\right)  +S_{k}^{\odot d}\left(  t\right)  \left(
C\otimes I\right)  \nonumber\\
&  +S_{k}\left(  t\right)  \left(  I\otimes T\right)  +S_{k+1}\left(
t\right)  \left(  I\otimes T^{0}\alpha\right)  .\label{Equ6}%
\end{align}

\begin{Rem}
(a) For the supermarket model, many papers, such as Mitzenmacher
\cite{Mit:1996a} and Luczak and McDiarmid \cite{Luc:2006}, assumed that the
arrival process is Poisson with rate $n\lambda$. As a direct generalization of
the Poisson arrivals with rate $n\lambda$, this paper uses a MAP with an
irreducible matrix descriptor $\left(  nC,nD\right)  $ of size $m_{A}$ whose
stationary arrival rate is given by $n\lambda=n\gamma De$.

(b) When there are $n$ servers in the supermarket model, we may use a more
general MAP with an irreducible matrix descriptor $\left(  C_{n},D_{n}\right)
$ of size $m_{A}$, where%
\[
\lim_{n\rightarrow\infty}\frac{C_{n}}{n}=C,\text{ \ }\lim_{n\rightarrow\infty
}\frac{D_{n}}{n}=D,
\]
and $\left(  C,D\right)  $ is also the irreducible matrix descriptor of a MAP.
It is easy to see from the above analysis that we can also obtain the system
of differential vector equations (\ref{Equ3}) to (\ref{Equ6}) with respect to
the more general MAP.
\end{Rem}

\section{Doubly Exponential Solution}

In this section, we provide a novel matrix-analytic approach for computing the
fixed point of the system of differential vector equations (\ref{Equ3}) to
(\ref{Equ6}), and give a closed-form solution with a doubly exponential
structure to the fixed point.

A row vector $\pi=\left(  \pi_{0},\pi_{1},\pi_{2},\ldots\right)  $ is called a
fixed point of the fraction vector $S\left(  t\right)  $ if $\lim
_{t\rightarrow+\infty}S\left(  t\right)  =\pi$. In this case, it is easy to
see that%
\[
\lim_{t\rightarrow+\infty}\left[  \frac{\mathtt{d}}{\text{d}t}S\left(
t\right)  \right]  =0.
\]
Therefore, as $t\rightarrow+\infty$ the system of differential vector
equations (\ref{Equ3}) to (\ref{Equ6}) can be simplified as a system of
nonlinear equations as follows:%
\begin{equation}
\pi_{0}e=1, \label{Equ7}%
\end{equation}%
\begin{equation}
\pi_{0}^{\odot d}C++\pi_{1}\left(  I\otimes T^{0}\right)  =0, \label{Equ8}%
\end{equation}%
\begin{equation}
\pi_{0}^{\odot d}\left(  D\otimes\alpha\right)  +\pi_{1}^{\odot d}\left(
C\otimes I\right)  +\pi_{1}\left(  I\otimes T\right)  +\pi_{2}\left(  I\otimes
T^{0}\alpha\right)  =0, \label{Equ9}%
\end{equation}
and for $k\geq2$,%
\begin{equation}
\pi_{k-1}^{\odot d}\left(  D\otimes I\right)  +\pi_{k}^{\odot d}\left(
C\otimes I\right)  +\pi_{k}\left(  I\otimes T\right)  +\pi_{k+1}\left(
I\otimes T^{0}\alpha\right)  =0. \label{Equ10}%
\end{equation}

It is very challenging to solve the system of nonlinear equations (\ref{Equ7})
to (\ref{Equ10}). Here, our goal is to derive a closed-form solution with a
doubly exponential structure to the fixed point $\pi=(\pi_{0},\pi_{1},\pi
_{2},...)$ through a novel matrix-analytic approach.

It follows from Equations (\ref{Equ9}) and (\ref{Equ10}) that
\begin{align}
&  \left(  \pi_{1}^{\odot d},\pi_{2}^{\odot d},\pi_{3}^{\odot d}%
,\ldots\right)  \left(
\begin{array}
[c]{ccccc}%
C\otimes I & D\otimes I &  &  & \\
& C\otimes I & D\otimes I &  & \\
&  & C\otimes I & D\otimes I & \\
&  &  & \ddots & \ddots
\end{array}
\right) \nonumber\\
+  &  \left(  \pi_{1},\pi_{2},\pi_{3},\ldots\right)  \left(
\begin{array}
[c]{cccc}%
I\otimes T &  &  & \\
I\otimes T^{0}\alpha & I\otimes T &  & \\
& I\otimes T^{0}\alpha & I\otimes T & \\
&  & \ddots & \ddots
\end{array}
\right) \nonumber\\
&  =-\left(  \pi_{0}^{\odot d}\left(  D\otimes\alpha\right)  ,0,0,\ldots
\right)  . \label{Equ11}%
\end{align}

Let%
\[
A=\left(
\begin{array}
[c]{cccc}%
I\otimes T &  &  & \\
I\otimes T^{0}\alpha & I\otimes T &  & \\
& I\otimes T^{0}\alpha & I\otimes T & \\
&  & \ddots & \ddots
\end{array}
\right)  .
\]
Then it is easy to check that the matrix $A$ is invertible, and%
\[
-A^{-1}=\left(
\begin{array}
[c]{ccccc}%
I\otimes\left(  -T\right)  ^{-1} &  &  &  & \\
I\otimes\left[  e\alpha\left(  -T\right)  ^{-1}\right]   & I\otimes\left(
-T\right)  ^{-1} &  &  & \\
I\otimes\left[  e\alpha\left(  -T\right)  ^{-1}\right]   & I\otimes\left[
e\alpha\left(  -T\right)  ^{-1}\right]   & I\otimes\left(  -T\right)  ^{-1} &
& \\
I\otimes\left[  e\alpha\left(  -T\right)  ^{-1}\right]   & I\otimes\left[
e\alpha\left(  -T\right)  ^{-1}\right]   & I\otimes\left[  e\alpha\left(
-T\right)  ^{-1}\right]   & I\otimes\left(  -T\right)  ^{-1} & \\
\vdots & \vdots & \vdots & \vdots & \ddots
\end{array}
\right)  ,
\]%
\[
\left(
\begin{array}
[c]{ccccc}%
C\otimes I & D\otimes I &  &  & \\
& C\otimes I & D\otimes I &  & \\
&  & C\otimes I & D\otimes I & \\
&  &  & \ddots & \ddots
\end{array}
\right)  \left(  -A^{-1}\right)  =\left(
\begin{array}
[c]{ccccc}%
R & V &  &  & \\
W & R & V &  & \\
W & W & R & V & \\
\vdots & \vdots & \vdots & \vdots & \ddots
\end{array}
\right)  ,
\]
and%
\[
\left(  \pi_{0}^{\odot d}\left(  D\otimes\alpha\right)  ,0,0,\ldots\right)
\left(  -A^{-1}\right)  =\left(  \pi_{0}^{\odot d}D\otimes\left[
\alpha\left(  -T\right)  ^{-1}\right]  ,0,0,\ldots\right)  ,
\]
where%
\[
V=D\otimes\left(  -T\right)  ^{-1},
\]%
\[
W=\left(  C+D\right)  \otimes\left[  e\alpha\left(  -T\right)  ^{-1}\right]
\]
and%
\[
R=C\otimes\left(  -T\right)  ^{-1}+D\otimes\left[  e\alpha\left(  -T\right)
^{-1}\right]  .
\]
Thus it follows from (\ref{Equ11}) that%
\begin{align}
\left(  \pi_{1},\pi_{2},\pi_{3},\ldots\right)  = &  \left(  \pi_{1}^{\odot
d},\pi_{2}^{\odot d},\pi_{3}^{\odot d},\ldots\right)  \left(
\begin{array}
[c]{ccccc}%
R & V &  &  & \\
W & R & V &  & \\
W & W & R & V & \\
\vdots & \vdots & \vdots & \vdots & \ddots
\end{array}
\right)  \nonumber\\
&  +\left(  \pi_{0}^{\odot d}D\otimes\left[  \alpha\left(  -T\right)
^{-1}\right]  ,0,0,\ldots\right)  ,\label{Equ12}%
\end{align}
which leads to a new system of nonlinear equations as follows:%
\begin{equation}
\pi_{1}=\pi_{0}^{\odot d}D\otimes\left[  \alpha\left(  -T\right)
^{-1}\right]  +\pi_{1}^{\odot d}R+\sum_{j=2}^{\infty}\pi_{j}^{\odot
d}W\label{Equ13}%
\end{equation}
and for $k\geq2$,%
\begin{equation}
\pi_{k}=\pi_{k-1}^{\odot d}V+\pi_{k}^{\odot d}R+\sum_{j=k+1}^{\infty}\pi
_{j}^{\odot d}W.\label{Equ14}%
\end{equation}
Now, we need to omit the two terms $\pi_{l}^{\odot d}R$ for $l\geq1$ and
$\sum_{j=k}^{\infty}\pi_{j}^{\odot d}W$ for $k\geq2$ in Equations
(\ref{Equ13}) and (\ref{Equ14}). Note that the Markov chain $C+D$ is positive
recurrent, we assume that the system of nonlinear equations (\ref{Equ13}) and
(\ref{Equ14}) has a closed-form solution%
\begin{equation}
\pi_{0}=\theta\gamma^{\odot\frac{1}{d}}\label{Equ14-1}%
\end{equation}
and for $k\geq1$%
\begin{equation}
\pi_{k}=r\left(  k\right)  \left(  \gamma^{\odot\frac{1}{d}}\otimes
\alpha^{\odot\frac{1}{d}}\right)  ,\label{Equ15}%
\end{equation}
where $\theta=1/\gamma^{\odot\frac{1}{d}}e$, and $r\left(  k\right)  $ is a
positive constant for $k\geq1$. Then it follows from (\ref{Equ13}),
(\ref{Equ14}) and (\ref{Equ15}) that%
\begin{align}
r\left(  1\right)  \left(  \gamma^{\odot\frac{1}{d}}\otimes\alpha
^{\odot\frac{1}{d}}\right)  = &  \pi_{0}^{\odot d}D\otimes\left[
\alpha\left(  -T\right)  ^{-1}\right]  +r^{d}\left(  1\right)  \left(
\gamma\otimes\alpha\right)  R\nonumber\\
&  +\sum_{j=2}^{\infty}r^{d}\left(  j\right)  \left(  \gamma\otimes
\alpha\right)  W;\label{Equ16}%
\end{align}
and for $k\geq2$,%
\begin{align}
r\left(  k\right)  \left(  \gamma^{\odot\frac{1}{d}}\otimes\alpha
^{\odot\frac{1}{d}}\right)  = &  r^{d}\left(  k-1\right)  \left(
\gamma\otimes\alpha\right)  V+r^{d}\left(  k\right)  \left(  \gamma
\otimes\alpha\right)  R\nonumber\\
&  +\sum_{j=k+1}^{\infty}r^{d}\left(  j\right)  \left(  \gamma\otimes
\alpha\right)  W.\label{Equ17}%
\end{align}
Note that%
\begin{align*}
\left(  \gamma\otimes\alpha\right)  W &  =\left(  \gamma\otimes\alpha\right)
\left\{  \left(  C+D\right)  \otimes\left[  e\alpha\left(  -T\right)
^{-1}\right]  \right\}  \\
&  =\gamma\left(  C+D\right)  \otimes\alpha\left[  e\alpha\left(  -T\right)
^{-1}\right]
\end{align*}
and%
\[
\gamma\left(  C+D\right)  =0,
\]
it is clear that%
\[
\left(  \gamma\otimes\alpha\right)  W=0.
\]
Similarly, we can compute%
\begin{align*}
\left(  \gamma\otimes\alpha\right)  R &  =\left(  \gamma\otimes\alpha\right)
\left\{  C\otimes\left(  -T\right)  ^{-1}+D\otimes\left[  e\alpha\left(
-T\right)  ^{-1}\right]  \right\}  \\
&  =\gamma\left(  C+D\right)  \otimes\alpha\left(  -T\right)  ^{-1}=0.
\end{align*}
It follows from (\ref{Equ16}) and (\ref{Equ17}) that%
\begin{equation}
\pi_{1}=\pi_{0}^{\odot d}D\otimes\left[  \alpha\left(  -T\right)
^{-1}\right]  \label{Equ17-1}%
\end{equation}
or%
\begin{equation}
r\left(  1\right)  \left(  \gamma^{\odot\frac{1}{d}}\otimes\alpha
^{\odot\frac{1}{d}}\right)  =\pi_{0}^{\odot d}D\otimes\left[  \alpha\left(
-T\right)  ^{-1}\right]  ;\label{Equ18}%
\end{equation}
and for $k\geq2$%
\begin{equation}
\pi_{k}=\pi_{k-1}^{\odot d}\left(  D\otimes\left(  -T\right)  ^{-1}\right)
\label{Equ18-1}%
\end{equation}
or%
\begin{equation}
r\left(  k\right)  \left(  \gamma^{\odot\frac{1}{d}}\otimes\alpha
^{\odot\frac{1}{d}}\right)  =r^{d}\left(  k-1\right)  \left(  \gamma
\otimes\alpha\right)  \left(  D\otimes\left(  -T\right)  ^{-1}\right)
.\label{Equ19}%
\end{equation}
Let $\omega=1/\alpha^{\odot\frac{1}{d}}e$. Then $0<\theta,\omega<1$ due to
$\gamma^{\odot\frac{1}{d}}e>1$ and $\alpha^{\odot\frac{1}{d}}e>1$. Note that
$\pi_{0}=\theta\gamma^{\odot\frac{1}{d}}$, $\lambda=\gamma De$, $1/\mu
=\alpha\left(  -T\right)  ^{-1}e$ and $\rho=\lambda/\mu$, it follows from
(\ref{Equ18}) and (\ref{Equ19}) that%
\begin{equation}
r\left(  1\right)  =\theta^{d}\left(  \theta\omega\rho\right)  \label{Equ20}%
\end{equation}
and for $k\geq2$%
\begin{align}
r\left(  k\right)   &  =r^{d}\left(  k-1\right)  \theta\omega\rho\nonumber\\
&  =\left[  r\left(  1\right)  \right]  ^{d^{k-1}}\left(  \theta\omega
\rho\right)  ^{d^{k-2}+d^{k-3}+\cdots+1}\nonumber\\
&  =\theta^{d^{k}}\left(  \theta\omega\rho\right)  ^{d^{k-1}+d^{k-2}+\cdots
+1}\nonumber\\
&  =\theta^{d^{k}}\left(  \theta\omega\rho\right)  ^{\frac{d^{k}-1}{d-1}%
}.\label{Equ21}%
\end{align}
It is easy to see from (\ref{Equ14-1}), (\ref{Equ15}) and (\ref{Equ21}) that%
\begin{equation}
\pi_{0}=\theta\cdot\gamma^{\odot\frac{1}{d}}\label{Equ22}%
\end{equation}
and for $k\geq1$%
\begin{equation}
\pi_{k}=\theta^{d^{k}}\left(  \theta\omega\rho\right)  ^{\frac{d^{k}-1}{d-1}%
}\cdot\gamma^{\odot\frac{1}{d}}\otimes\alpha^{\odot\frac{1}{d}}.\label{Equ23}%
\end{equation}
Now, we use (\ref{Equ22}) and (\ref{Equ23}) to check Equations (\ref{Equ8})
and (\ref{Equ17-1}) that%
\[
\left\{
\begin{array}
[c]{l}%
\pi_{0}^{\odot d}C+\pi_{1}\left(  I\otimes T^{0}\right)  =0,\\
\pi_{1}=\pi_{0}^{\odot d}D\otimes\left[  \alpha\left(  -T\right)
^{-1}\right]  ,
\end{array}
\right.
\]
which leads to%
\begin{equation}
\pi_{0}^{\odot d}\left(  C+D\right)  =0.\label{Equ24}%
\end{equation}
Obviously, $\pi_{0}=\theta\cdot\gamma^{\odot\frac{1}{d}}$ is a non-zero
nonnegative solution to Equation (\ref{Equ24}), and $\pi_{0}e=1$.

Summarizing the above analysis, the following theorem describes a closed-form
solution with a doubly exponential structure to the fixed point.

\begin{The}
\label{The:FixedP}The fixed point $\pi=\left(  \pi_{0},\pi_{1},\pi_{2}%
,\ldots\right)  $ is given by%
\[
\pi_{0}=\theta\cdot\gamma^{\odot\frac{1}{d}},
\]
and for $k\geq1,$%
\[
\pi_{k}=\theta^{d^{k}}\left(  \theta\omega\rho\right)  ^{\frac{d^{k}-1}{d-1}%
}\cdot\gamma^{\odot\frac{1}{d}}\otimes\alpha^{\odot\frac{1}{d}}.
\]
\end{The}

The following corollary indicates that the fixed point can be decomposed into
the product of two factors inflecting arrival information and service
information. Based on this, it is easy to see the role played by the arrival
and service processes in the fixed point.

\begin{Cor}
The fixed point $\pi=\left(  \pi_{0},\pi_{1},\pi_{2},\ldots\right)  $ can be
decomposed into the product of two factors inflecting arrival information and
service information
\[
\pi_{k}=\left\{  \theta^{\frac{d^{k+1}-1}{d-1}}\lambda^{\frac{d^{k}-1}{d-1}%
}\cdot\gamma^{\odot\frac{1}{d}}\right\}  \otimes\left\{  \left(  \frac{\omega
}{\mu}\right)  ^{\frac{d^{k}-1}{d-1}}\cdot\alpha^{\odot\frac{1}{d}}\right\}
,\text{ \ }k\geq0.
\]
\end{Cor}

\begin{Rem}
We consider a supermarket model with Poisson arrivals with rate $\lambda$ and
exponential service times with rate $\mu$, which has been extensively analyzed
in the literature. Obviously, $\pi_{0}=1$. It follows from (\ref{Equ11}) that%
\begin{align*}
&  \left(  \pi_{1}^{\odot d},\pi_{2}^{\odot d},\pi_{3}^{\odot d}%
,\ldots\right)  \left(
\begin{array}
[c]{ccccc}%
-\lambda & \lambda &  &  & \\
& -\lambda & \lambda &  & \\
&  & -\lambda & \lambda & \\
&  &  & \ddots & \ddots
\end{array}
\right)  \newline \\
+ &  \left(  \pi_{1},\pi_{2},\pi_{3},\ldots\right)  \left(
\begin{array}
[c]{cccc}%
-\mu &  &  & \\
\mu & -\mu &  & \\
& \mu & -\mu & \\
&  & \ddots & \ddots
\end{array}
\right)  =-\left(  \lambda,0,0,\ldots\right)  ,
\end{align*}
which leads to%
\[
\left(  \pi_{1},\pi_{2},\pi_{3},\ldots\right)  =\left(  \pi_{1}^{\odot d}%
,\pi_{2}^{\odot d},\pi_{3}^{\odot d},\ldots\right)  \left(
\begin{array}
[c]{ccccc}%
0 & \rho &  &  & \\
& 0 & \rho &  & \\
&  & 0 & \rho & \\
&  &  & \ddots & \ddots
\end{array}
\right)  +\left(  \rho,0,0,\ldots\right)  .
\]
Thus we obtain%
\[
\pi_{1}=\rho
\]
and for $k\geq2$,%
\[
\pi_{k}=\rho\pi_{k-1}^{\odot d}=\rho^{d^{k-1}+d^{k-2}+\cdots+1}=\rho
^{\frac{d^{k}-1}{d-1}},
\]
which is the same as Lemma 3.2 in Mitzenmacher \cite{Mit:1996b}.
\end{Rem}

Based on Theorem \ref{The:FixedP}, we now compute the expected sojourn time
$T_{d}$\ that a tagged arriving customer spends in the supermarket model. For
the PH service time $X$ with an irreducible representation $\left(
\alpha,T\right)  $, the residual time $X_{R}$ of $X$ is also of phase type
with an irreducible representation $\left(  \tau,T\right)  $, where $\tau$ is
the stationary probability vector of the Markov chain $T+T^{0}\alpha$. Thus,
we have%
\[
E\left[  X\right]  =\alpha\left(  -T\right)  ^{-1}e=\frac{1}{\mu},\text{
\ }E\left[  X_{R}\right]  =\tau\left(  -T\right)  ^{-1}e.
\]

For the PH service times, a tagged arriving customer is the $k$th customer in
the corresponding queue with probability $\pi_{k-1}^{\odot d}e-\pi_{k}^{\odot
d}e$. Thus it is easy to see that the expected sojourn time of the tagged
arriving customer is given by%
\begin{align*}
E\left[  T_{d}\right]   &  =\left(  \pi_{0}^{\odot d}e-\pi_{1}^{\odot
d}e\right)  E\left[  X\right]  +\sum_{k=1}^{\infty}\left(  \pi_{k}^{\odot
d}e-\pi_{k+1}^{\odot d}e\right)  \left[  E\left[  X_{R}\right]  +kE\left[
X\right]  \right]  \\
&  =\left\{  E\left[  X_{R}\right]  -E\left[  X\right]  \right\}  \pi
_{1}^{\odot d}e+E\left[  X\right]  \sum_{k=0}^{\infty}\pi_{k}^{\odot d}e\\
&  =\theta^{d^{2}}\left(  \theta\omega\rho\right)  ^{d}\left(  \tau
-\alpha\right)  \left(  -T\right)  ^{-1}e+\frac{1}{\mu}\sum_{k=0}^{\infty
}\theta^{d^{k+1}}\left(  \theta\omega\rho\right)  ^{\frac{d^{k+1}-d}{d-1}}.
\end{align*}
When the arrival process and the service time distribution are Poisson and
exponential, respectively, it is clear that $\alpha=\tau=1$ and $\theta
=\omega=1$, thus we have%
\[
E\left[  T_{d}\right]  =\frac{1}{\mu}\sum_{k=0}^{\infty}\rho^{\frac{d^{k+1}%
-d}{d-1}},
\]
which is the same as Corollary 3.8 in Mitzenmacher \cite{Mit:1996b}.

In what follows we provide an example to indicate how the expected sojourn
time $E[T_{d}]$ depends on the choice number $d$. We assume that $m=2$ and
\[
C=\left(
\begin{array}
[c]{cc}%
-10 & 7\\
4 & -9
\end{array}
\right)  ,\ \ D=\left(
\begin{array}
[c]{cc}%
1 & 2\\
3 & 2
\end{array}
\right)  ,
\]
and the service times are exponential with service rate $\mu=5,10,20$,
respectively. It is seen from Figure 2 that the expected sojourn time
$E[T_{d}]$ decreases very fast as the choice number $d$ increases.

\begin{figure}[tbh]
\centering                        \includegraphics[width=10cm]{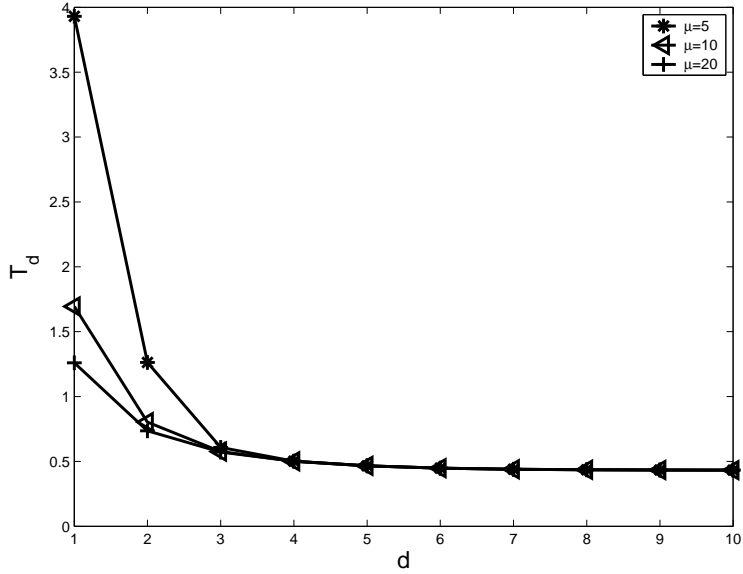}
\caption{$E[T_{d}]$ vs $d$ for the MAP}%
\end{figure}

\section{An Important Observation}

In this section, we analyze a special supermarket model with Poisson arrivals
an PH service times, and obtain two different doubly exponential solutions to
the fixed point. Based on this, we give an important observation, namely that
the doubly exponential solution to the fixed point is not always unique for
more general supermarket models.

When the arrival process is Poisson, it follows from (\ref{Equ7}) to
(\ref{Equ10}) that%
\begin{equation}
\pi_{0}=1, \label{Equ25}%
\end{equation}%
\begin{equation}
-\lambda+\pi_{1}T^{0}=0, \label{Equ26}%
\end{equation}%
\begin{equation}
\lambda\alpha-\lambda\pi_{1}^{\odot d}+\pi_{1}T+\pi_{2}T^{0}\alpha=0,
\label{Equ27}%
\end{equation}
and for $k\geq2$,%
\begin{equation}
\lambda\pi_{k-1}^{\odot d}\left(  t\right)  -\lambda\pi_{k}^{\odot d}\left(
t\right)  +\pi_{k}\left(  t\right)  T+\pi_{k+1}\left(  t\right)  T^{0}%
\alpha=0. \label{Equ28}%
\end{equation}

For the system of nonlinear equations (\ref{Equ25}) to (\ref{Equ28}), we can
provide two different doubly exponential solutions to the fixed point
$\pi=\left(  \pi_{0},\pi_{1},\pi_{2},\ldots\right)  $.

\subsection{The first doubly exponential solution}

The first doubly exponential solution has been given in Section 3. Here, we
simply list the crucial derivations for the special supermarket model.

It follows from (\ref{Equ11}) that
\begin{align*}
&  \left(  \pi_{1}^{\odot d},\pi_{2}^{\odot d},\pi_{3}^{\odot d}%
,\ldots\right)  \left(
\begin{array}
[c]{ccccc}%
-\lambda & \lambda &  &  & \\
& -\lambda & \lambda &  & \\
&  & -\lambda & \lambda & \\
&  &  & \ddots & \ddots
\end{array}
\right)  +\left(  \pi_{1},\pi_{2},\pi_{3},\ldots\right)  \left(
\begin{array}
[c]{cccc}%
T &  &  & \\
T^{0}\alpha & T &  & \\
& T^{0}\alpha & T & \\
&  & \ddots & \ddots
\end{array}
\right) \\
&  =-\left(  \lambda\alpha,0,0,\ldots\right)  ,
\end{align*}
which leads to%
\begin{align*}
\left(  \pi_{1},\pi_{2},\pi_{3},\ldots\right)  =  &  \left(  \pi_{1}^{\odot
d},\pi_{2}^{\odot d},\pi_{3}^{\odot d},\ldots\right)  \left(
\begin{array}
[c]{ccccc}%
R & V &  &  & \\
& R & V &  & \\
&  & R & V & \\
&  &  & \ddots & \ddots
\end{array}
\right) \\
&  +\left(  \lambda\alpha\left(  -T\right)  ^{-1},0,0,\ldots\right)  ,
\end{align*}
where%
\[
V=\lambda\left(  -T\right)  ^{-1}%
\]
and%
\[
R=\lambda\left(  -I+e\alpha\right)  \left(  -T\right)  ^{-1}.
\]
Thus we obtain%
\begin{equation}
\pi_{1}=\lambda\alpha\left(  -T\right)  ^{-1}+\pi_{1}^{\odot d}\left[
\lambda\left(  -I+e\alpha\right)  \left(  -T\right)  ^{-1}\right]
\label{Equ29}%
\end{equation}
and for $k\geq2$%
\begin{equation}
\pi_{k}=\pi_{k-1}^{\odot d}\left[  \lambda\left(  -T\right)  ^{-1}\right]
+\pi_{k}^{\odot d}\left[  \lambda\left(  -I+e\alpha\right)  \left(  -T\right)
^{-1}\right]  . \label{Equ30}%
\end{equation}
To omit the term $\pi_{k}^{\odot d}\left[  \lambda\left(  -I+e\alpha\right)
\left(  -T\right)  ^{-1}\right]  $ for $k\geq1$, we assume that $\left\{
\pi_{k},k\geq1\right\}  $ has the following expression%
\[
\pi_{k}=r\left(  k\right)  \alpha^{\odot\frac{1}{d}}.
\]
In this case, we have%
\[
\pi_{k}^{\odot d}\left[  \lambda\left(  -I+e\alpha\right)  \left(  -T\right)
^{-1}\right]  =r^{d}\left(  k\right)  \alpha\left[  \lambda\left(
-I+e\alpha\right)  \left(  -T\right)  ^{-1}\right]  =0,
\]
thus it follows from (\ref{Equ29}) and (\ref{Equ30}) that%
\begin{equation}
\pi_{1}=\lambda\alpha\left(  -T\right)  ^{-1} \label{Equ31}%
\end{equation}
and for $k\geq2$%
\begin{equation}
\pi_{k}=\pi_{k-1}^{\odot d}\left[  \lambda\left(  -T\right)  ^{-1}\right]  .
\label{Equ32}%
\end{equation}
It follows from (\ref{Equ31}) that%
\[
r\left(  1\right)  \alpha^{\odot\frac{1}{d}}=\lambda\alpha\left(  -T\right)
^{-1},
\]
which follows that%
\[
r\left(  1\right)  =\omega\rho.
\]
It follows from (\ref{Equ32}) that%
\[
r\left(  k\right)  \alpha^{\odot\frac{1}{d}}=r^{d}\left(  k-1\right)
\alpha\left[  \lambda\left(  -T\right)  ^{-1}\right]
\]
which follows that%
\[
r\left(  k\right)  =r^{d}\left(  k-1\right)  \omega\rho=\left(  \omega
\rho\right)  ^{\frac{d^{k}-1}{d-1}}.
\]
Therefore, we can obtain%
\[
\pi_{0}=1
\]
and for $k\geq1$%
\begin{equation}
\pi_{k}=\left(  \omega\rho\right)  ^{\frac{d^{k}-1}{d-1}}\cdot\alpha
^{\odot\frac{1}{d}}. \label{Equ33}%
\end{equation}

\subsection{The second doubly exponential solution}

The second doubly exponential solution was given in Li, Lui and Wang
\cite{Li:2010a}, thus we provide some crucial computational steps.

It follows from (\ref{Equ26}) that%
\begin{equation}
\pi_{1}T^{0}=\lambda. \label{Equ34}%
\end{equation}
To solve Equation (\ref{Equ34}), we denote by $\tau$ the stationary
probability vector of the irreducible Markov chain $T+T^{0}\alpha$. Obviously,
we have%
\[
\tau T^{0}=\mu,
\]%
\begin{equation}
\frac{\lambda}{\mu}\tau T^{0}=\lambda. \label{Equ35}%
\end{equation}
Thus, we obtain%
\[
\pi_{1}=\frac{\lambda}{\mu}\tau=\rho\cdot\tau.
\]
Using $\pi_{0}=1$ and $\pi_{1}=\rho\cdot\tau$, it follows from Equation
(\ref{Equ27}) that%
\[
\lambda\alpha-\lambda\rho^{d}\cdot\tau^{\odot d}+\rho\cdot\tau T+\pi_{2}%
T^{0}\alpha=0,
\]
which leads to%
\[
\lambda-\lambda\rho^{d}\cdot\tau^{\odot d}e+\rho\cdot\tau Te+\pi_{2}T^{0}=0.
\]
Note that $\tau Te=-\mu$ and $\rho=\lambda/\mu$, we obtain%
\[
\pi_{2}T^{0}=\lambda\rho^{d}\tau^{\odot d}e.
\]
Let $\psi=\tau^{\odot d}e$. Then it is easy to see that $\psi\in\left(
0,1\right)  $, and%
\[
\pi_{2}T^{0}=\lambda\psi\rho^{d}.
\]
Using similar analysis on Equation (\ref{Equ34}), we have%
\[
\pi_{2}=\frac{\lambda\psi\rho^{d}}{\mu}\tau=\psi\rho^{d+1}\cdot\tau.
\]
Based on $\pi_{1}=\rho\cdot\tau$ and $\pi_{2}=\psi\rho^{d+1}\cdot\tau$, it
follows from Equation (\ref{Equ28}) that for $k=2$,
\[
\lambda\rho^{d}\cdot\tau^{\odot d}-\lambda\psi^{d}\rho^{d^{2}+d}\cdot
\tau^{\odot d}+\psi\rho^{d+1}\cdot\tau T+\pi_{3}T^{0}\alpha=0,
\]
which leads to%
\[
\lambda\psi\rho^{d}-\lambda\psi^{d+1}\rho^{d^{2}+d}+\psi\rho^{d+1}\cdot\tau
Te+\pi_{3}T^{0}=0,
\]
thus we obtain%
\[
\pi_{3}T^{0}=\lambda\psi^{d+1}\rho^{d^{2}+d}.
\]
Using a similar analysis on Equation (\ref{Equ34}), we have%
\[
\pi_{3}=\frac{\lambda\psi^{d+1}\rho^{d^{2}+d}}{\mu}\tau=\psi^{d+1}\rho
^{d^{2}+d+1}\cdot\tau.
\]
Now, we assume that $\pi_{k}=\psi^{\frac{d^{k-1}-1}{d-1}}\rho^{\frac{d^{k}%
-1}{d-1}}\cdot\tau$ is correct for the cases with $l=k$. Then for $l=k+1$ we
have
\begin{align*}
\lambda &  \psi^{d^{k-2}+d^{k-3}+\cdots+d}\rho^{d^{k-1}+d^{k-2}+\cdots+d}%
\cdot\tau^{\odot d}-\lambda\psi^{d^{k-1}+d^{k-2}+\cdots+d}\rho^{d^{k}%
+d^{k-1}+\cdots+d}\cdot\tau^{\odot d}\\
&  +\psi^{d^{k-2}+d^{k-3}+\cdots+1}\rho^{d^{k-1}+d^{k-2}+\cdots+1}\cdot\tau
T+\pi_{k+1}T^{0}\alpha=0,
\end{align*}
which leads to%
\begin{align*}
\lambda &  \psi^{d^{k-2}+d^{k-3}+\cdots+d+1}\rho^{d^{k-1}+d^{k-2}+\cdots
+d}-\lambda\psi^{d^{k-1}+d^{k-2}+\cdots+d+1}\rho^{d^{k}+d^{k-1}+\cdots+d}\\
&  +\psi^{d^{k-2}+d^{k-3}+\cdots+1}\rho^{d^{k-1}+d^{k-2}+\cdots+1}\cdot\tau
Te+\pi_{k+1}T^{0}=0,
\end{align*}
thus we obtain%
\[
\pi_{k+1}T^{0}=\lambda\psi^{d^{k-1}+d^{k-2}+\cdots+d+1}\rho^{d^{k}%
+d^{k-1}+\cdots+d}.
\]
By a similar analysis to (\ref{Equ34}) and (\ref{Equ35}), we have%
\begin{align*}
\pi_{k+1}  &  =\frac{\lambda\psi^{d^{k-1}+d^{k-2}+\cdots+d+1}\rho
^{d^{k}+d^{k-1}+\cdots+d}}{\mu}\tau\\
&  =\psi^{d^{k-1}+d^{k-2}+\cdots+d+1}\rho^{d^{k}+d^{k-1}+\cdots+d+1}\cdot\tau.
\end{align*}
Therefore, by induction we can obtain%
\[
\pi_{0}=1,
\]
and for $k\geq1$%
\begin{equation}
\pi_{k}=\psi^{\frac{d^{k-1}-1}{d-1}}\rho^{\frac{d^{k}-1}{d-1}}\cdot\tau.
\label{Equ36}%
\end{equation}

\subsection{An important observation}

Now, we have given two expressions (\ref{Equ33}) and (\ref{Equ36}) for the
fixed point. In what follows we provide some examples to indicate that the two
expressions may be different from each other.

\textbf{Example one: }When the PH service time is exponential, it is easy to
see that $\alpha=\tau=1$, which leads to that $\omega=\psi=1$. Thus the fixed
point is given by%
\[
\pi_{k}=\rho^{\frac{d^{k}-1}{d-1}},k\geq1.
\]
In this case, (\ref{Equ33}) is the same as (\ref{Equ36}).

\textbf{Example two: }When the service time is an $m$-order Erlang
distribution with an irreducible representation $(\alpha,T)$, where%
\[
\alpha=\left(  1,0,\ldots,0\right)
\]
and%
\[
T=\left(
\begin{array}
[c]{ccccc}%
-\eta & \eta &  &  & \\
& -\eta & \eta &  & \\
&  & \ddots & \ddots & \\
&  &  & -\eta & \eta\\
&  &  &  & -\eta
\end{array}
\right)  ,\text{ \ \ }T^{0}=\left(
\begin{array}
[c]{c}%
0\\
0\\
\vdots\\
0\\
\eta
\end{array}
\right)  .
\]
We have%
\[
\alpha^{\odot\frac{1}{d}}=\left(  1,0,\ldots,0\right)
\]
and%
\[
\omega=\frac{1}{\alpha^{\odot\frac{1}{d}}e}=1.
\]
Thus the first doubly exponential solution is given by%
\begin{equation}
\pi_{k}^{\left(  \text{F}\right)  }=\rho^{\frac{d^{k}-1}{d-1}}\cdot\left(
1,0,\ldots,0\right)  ,\text{ \ }k\geq1.\label{Equ37}%
\end{equation}
It is clear that%
\[
T+T^{0}\alpha=\left(
\begin{array}
[c]{ccccc}%
-\eta & \eta &  &  & \\
& -\eta & \eta &  & \\
&  & \ddots & \ddots & \\
&  &  & -\eta & \eta\\
\eta &  &  &  & -\eta
\end{array}
\right)  ,
\]
which leads to the stationary probability vector of the Markov chain
$T+T^{0}\alpha$ as follows:
\[
\tau=\left(  \frac{1}{m},\frac{1}{m},\ldots,\frac{1}{m}\right)  ,
\]%
\[
\mu=\tau T^{0}=\frac{\eta}{m},
\]%
\[
\rho=\frac{\lambda}{\mu}=\frac{m\lambda}{\eta}%
\]
and%
\[
\psi=m\left(  \frac{1}{m}\right)  ^{d}=m^{1-d}.
\]
Thus the second doubly exponential solution is given by
\begin{align}
\pi_{k}^{\left(  \text{S}\right)  } &  =\psi^{\frac{d^{k-1}-1}{d-1}}%
\rho^{\frac{d^{k}-1}{d-1}}\left(  \frac{1}{m},\frac{1}{m},\ldots,\frac{1}%
{m}\right)  \nonumber\\
&  =\rho^{\frac{d^{k}-1}{d-1}}\cdot\left(  m^{-d^{k-1}},m^{-d^{k-1}}%
,\ldots,m^{-d^{k-1}}\right)  ,\text{ \ }k\geq1.\label{Equ38}%
\end{align}
It is clear that (\ref{Equ37}) and (\ref{Equ38}) are different from each other
for $k,m,d\geq2$, and%
\[
\frac{\pi_{k}^{\left(  \text{F}\right)  }e}{\pi_{k}^{\left(  \text{S}\right)
}e}=m^{d^{k-1}-1}.
\]
It is clear that $\pi_{k}^{\left(  \text{F}\right)  }e\neq\pi_{k}^{\left(
\text{S}\right)  }e$ for $k,m,d\geq2$.

\begin{Rem}
For the supermarket model with Poisson arrivals and PH service times, we have
obtained two different doubly exponential solutions to the fixed point. It is
interesting but difficult to be able to find another new doubly exponential
solution to the fixed point. Furthermore, we believe that it is an open
problem how to give all doubly exponential solutions to the fixed point for
more general supermarket models.
\end{Rem}

\section{Exponential Convergence}

In this section, we provide an upper bound for the current location $S\left(
t\right)  $ of the supermarket model, and study exponential convergence of the
current location $S\left(  t\right)  $ to its fixed point $\pi$.

For the supermarket model, the initial point $S\left(  0\right)  $ can affect
the current location $S\left(  t\right)  $ for each $t>0$, since the arrival
and service processes are under a unified structure through a sample path
comparison. To explain this, it is necessary to provide some notation for
comparison of two vectors. Let $a=\left(  a_{1},a_{2},a_{3},\ldots\right)  $
and $b=\left(  b_{1},b_{2},b_{3},\ldots\right)  $. We write $a\prec b$ if
$a_{k}<b_{k}$ for some $k\geq1$ and $a_{l}\leq b_{l}$ for $l\neq k,l\geq1$;
and $a\preceq b$ if $a_{k}\leq b_{k}$ for all $k\geq1$.

Now, we can easily obtain the following useful proposition, while the proof is
clear by means of a sample path analysis, and thus is omitted here.

\begin{Pro}
\label{Prop1}If $S\left(  0\right)  \preceq\widetilde{S}\left(  0\right)  $,
then $S\left(  t\right)  \preceq\widetilde{S}\left(  t\right)  $ for $t>0$.
\end{Pro}

Based on Proposition \ref{Prop1}, the following theorem shows that the fixed
point $\pi$ is an upper bound of the current location $S\left(  t\right)  $
for all $t\geq0$.

\begin{The}
\label{The:Bound}For the supermarket model, if there exists some $k$ such that
$S_{k}\left(  0\right)  =0$, then the sequence $\left\{  S_{k}\left(
t\right)  \right\}  $ for all $t\geq0$ has an upper bound sequence $\left\{
\pi_{k}\right\}  $ which decreases doubly exponentially, that is, $S\left(
t\right)  \preceq\pi$ for all $t\geq0$.
\end{The}

\noindent\textbf{Proof:} \ Let%
\[
\widetilde{S}_{k}\left(  0\right)  =\pi_{k},\text{ \ }k\geq0.
\]
Then for each $k\geq0$, $\widetilde{S}_{k}\left(  t\right)  =\widetilde{S}%
_{k}\left(  0\right)  =\pi_{k}$ for all $t\geq0$, since%
\[
\widetilde{S}\left(  0\right)  =\left(  \widetilde{S}_{0}\left(  0\right)
,\widetilde{S}_{1}\left(  0\right)  ,\widetilde{S}_{2}\left(  0\right)
,\ldots\right)  =\pi
\]
is a fixed point for the supermarket model. If $S_{k}\left(  0\right)  =0$ for
some $k$, then $S_{k}\left(  0\right)  \prec\widetilde{S}_{k}\left(  0\right)
$. Again, if $S_{j}\left(  0\right)  \preceq\widetilde{S}_{j}\left(  0\right)
$\ for all $j\neq k$, then $S\left(  0\right)  \preceq\widetilde{S}\left(
0\right)  $. It is easy to see from Proposition \ref{Prop1} that $S_{k}\left(
t\right)  \preceq\widetilde{S}_{k}\left(  t\right)  =\pi_{k}$ for all $k\geq0$
and $t\geq0$. Thus we obtain that for all $k\geq0$ and $t\geq0$
\[
S_{k}\left(  t\right)  \leq\pi_{k}=\theta^{d^{k}}\left(  \theta\omega
\rho\right)  ^{\frac{d^{k}-1}{d-1}}\cdot\gamma^{\odot\frac{1}{d}}\otimes
\alpha^{\odot\frac{1}{d}}.
\]
Since $0<\theta,\omega,\rho<1$, $\left\{  \pi_{k}\right\}  $ decreases doubly
exponentially. This completes the proof. \hspace*{\fill} \rule{1.8mm}{2.5mm}

To show exponential convergence, we define a Lyapunov function $\Phi\left(
t\right)  $ as%
\begin{equation}
\Phi\left(  t\right)  =\sum_{k=0}^{\infty}w_{k}\left(  \pi_{k}-S_{k}\left(
t\right)  \right)  e, \label{Equ39}%
\end{equation}
where $\left\{  w_{k}\right\}  $ is a positive scalar sequence with $w_{k}\geq
w_{k-1}\geq w_{0}=1$ for $k\geq2$.

The following theorem measures the distance of the current location $S\left(
t\right)  $ to the fixed point $\pi$ for $t\geq0$, and illustrates that this
distance will quickly come close to zero with exponential convergence. Hence,
it shows that from any suitable starting point, the supermarket model can be
quickly close to the fixed point, that is, there always exists a fixed point
in the supermarket model.

\begin{The}
For $t\geq0$,%
\[
\Phi\left(  t\right)  \leq c_{0}e^{-\delta t},
\]
where $c_{0}$ and $\delta$ are two positive constants. In this case, the
Lyapunov function $\Phi\left(  t\right)  $ is exponentially convergent.
\end{The}

\noindent\textbf{Proof:} \ It is seen from (\ref{Equ39}) that
\[
\frac{d}{dt}\Phi\left(  t\right)  =-\sum_{k=0}^{\infty}w_{k}\frac{d}{dt}%
S_{k}\left(  t\right)  e.
\]
It follows from Equations (\ref{Equ3}) to (\ref{Equ6}) that%
\begin{align*}
\frac{d}{dt}\Phi\left(  t\right)  =  &  -w_{0}\left[  S_{0}^{\odot d}\left(
t\right)  C+S_{1}\left(  t\right)  \left(  I\otimes T^{0}\right)  \right]  e\\
&  -w_{1}[S_{0}^{\odot d}\left(  t\right)  \left(  D\otimes\alpha\right)
+S_{1}^{\odot d}\left(  t\right)  \left(  C\otimes I\right) \\
&  +S_{1}\left(  t\right)  \left(  I\otimes T\right)  +S_{2}\left(  t\right)
\left(  I\otimes T^{0}\alpha\right)  ]e\\
&  -\sum_{k=2}^{\infty}w_{k}[S_{k-1}^{\odot d}\left(  t\right)  \left(
D\otimes I\right)  +S_{k}^{\odot d}\left(  t\right)  \left(  C\otimes I\right)
\\
&  +S_{k}\left(  t\right)  \left(  I\otimes T\right)  +S_{k+1}\left(
t\right)  \left(  I\otimes T^{0}\alpha\right)  ]e.
\end{align*}
By means of $Ce=-De$ and $Te=-T^{0}$, we can obtain%
\begin{align}
\frac{d}{dt}\Phi\left(  t\right)  =  &  -w_{0}\left[  S_{0}^{\odot d}\left(
t\right)  \left(  -De\right)  +S_{1}\left(  t\right)  \left(  e\otimes
T^{0}\right)  \right] \nonumber\\
&  -w_{1}[S_{0}^{\odot d}\left(  t\right)  \left(  De\right)  +S_{1}^{\odot
d}\left(  t\right)  \left(  \left(  -De\right)  \otimes e\right) \nonumber\\
&  +S_{1}\left(  t\right)  \left(  e\otimes\left(  -T^{0}\right)  \right)
+S_{2}\left(  t\right)  \left(  e\otimes T^{0}\right)  ]\nonumber\\
&  -\sum_{k=2}^{\infty}w_{k}[S_{k-1}^{\odot d}\left(  t\right)  \left(
\left(  De\right)  \otimes e\right)  +S_{k}^{\odot d}\left(  t\right)  \left(
\left(  -De\right)  \otimes e\right) \nonumber\\
&  +S_{k}\left(  t\right)  \left(  e\otimes\left(  -T^{0}\right)  \right)
+S_{k+1}\left(  t\right)  \left(  e\otimes T^{0}\right)  ]. \label{Equ40}%
\end{align}
Let%
\[
S_{0}^{\odot d}\left(  t\right)  \left(  De\right)  =c_{0}\left(  t\right)
\cdot\left[  \pi_{0}-S_{0}\left(  t\right)  \right]  e,
\]
for $k\geq1$%
\[
S_{k}^{\odot d}\left(  t\right)  \left(  \left(  De\right)  \otimes e\right)
=c_{k}\left(  t\right)  \cdot\left[  \pi_{k}-S_{k}\left(  t\right)  \right]
e
\]
and%
\[
S_{k}\left(  t\right)  \left(  e\otimes T^{0}\right)  =d_{k}\left(  t\right)
\cdot\left[  \pi_{k}-S_{k}\left(  t\right)  \right]  e.
\]
Then it follows from (\ref{Equ40}) that%
\begin{align*}
\frac{d}{dt}\Phi\left(  t\right)   &  =-\left(  w_{1}-w_{0}\right)
c_{0}\left(  t\right)  \cdot\left[  \pi_{0}-S_{0}\left(  t\right)  \right]
e\\
&  -\sum_{k=1}^{\infty}\left[  w_{k+1}c_{k}\left(  t\right)  -w_{k}\left(
c_{k}\left(  t\right)  +d_{k}\left(  t\right)  \right)  +w_{k-1}d_{k}\left(
t\right)  \right]  \cdot\left[  \pi_{k}-S_{k}\left(  t\right)  \right]  e.
\end{align*}
Let%
\[
w_{0}=1,
\]%
\[
\left(  w_{1}-w_{0}\right)  c_{0}\left(  t\right)  \geq\delta w_{0}%
\]
and%
\[
w_{k+1}c_{k}\left(  t\right)  -w_{k}\left(  c_{k}\left(  t\right)
+d_{k}\left(  t\right)  \right)  +w_{k-1}d_{k}\left(  t\right)  \geq\delta
w_{k}.
\]
Then%
\[
w_{1}\geq1+\frac{\delta}{c_{0}\left(  t\right)  },
\]%
\[
w_{2}\geq w_{1}+\frac{\delta w_{1}}{c_{1}\left(  t\right)  }+\frac{d_{1}%
\left(  t\right)  }{c_{1}\left(  t\right)  }\left(  w_{1}-1\right)
\]
and for $k\geq2$%
\[
w_{k+1}\geq w_{k}+\frac{\delta w_{k}}{c_{k}\left(  t\right)  }+\frac{d_{k}%
\left(  t\right)  }{c_{k}\left(  t\right)  }\left(  w_{k}-w_{k-1}\right)  .
\]
Thus we have
\[
\frac{d}{dt}\Phi\left(  t\right)  \leq-\delta\sum_{k=0}^{\infty}w_{k}\left[
\pi_{k}-S_{k}\left(  t\right)  \right]  e.
\]
It follows from (\ref{Equ39}) that%
\[
\frac{d}{dt}\Phi\left(  t\right)  \leq-\delta\Phi\left(  t\right)  ,
\]
which can leads to%
\[
\Phi\left(  t\right)  \leq c_{0}e^{-\delta t}.
\]
This completes the proof. \hspace*{\fill} \rule{1.8mm}{2.5mm}

\begin{Rem}
We have provided an algorithm for computing the positive scalar sequence
$\left\{  w_{k}\right\}  $ with $1=w_{0}\leq w_{k-1}<w_{k}$ for $k\geq2$ as follows:

Step one:%
\[
w_{0}=1.
\]

Step two:%
\[
w_{1}=1+\frac{\delta}{c_{0}\left(  t\right)  }%
\]
and%
\[
w_{2}=w_{1}+\frac{\delta w_{1}}{c_{1}\left(  t\right)  }+\frac{d_{1}\left(
t\right)  }{c_{1}\left(  t\right)  }\left(  w_{1}-1\right)
\]

Step three: for $k\geq2$%
\[
w_{k+1}=w_{k}+\frac{\delta w_{k}}{c_{k}\left(  t\right)  }+\frac{d_{k}\left(
t\right)  }{c_{k}\left(  t\right)  }\left(  w_{k}-w_{k-1}\right)  .
\]
This illustrates that $w_{k}$ is a function of time $t$ for $k\geq1$. Note
that $\lambda,\delta,c_{k}\left(  t\right)  ,d_{l}\left(  t\right)  >0$, it is
clear that for $k\geq2$%
\[
1=w_{0}\leq w_{k-1}<w_{k}.
\]
\end{Rem}

\section{A Lipschitz Condition}

In this section, we apply the Kurtz Theorem to study density dependent jump
Markov process given in the supermarket model with MAPs and PH service times,
which leads to the Lipschitz condition under which the fraction measure of the
supermarket model weakly converges the system of differential vector equations.

The supermarket model can be analyzed by a density dependent jump Markov
process, where the density dependent jump Markov process is a Markov process
with a single parameter $n$ which corresponds to the population size. Kurtz's
work provides a basis for density dependent jump Markov processes in order to
relate infinite-size systems of differential equations to corresponding
finite-size systems of differential equations. Readers may refer to Kurtz
\cite{Kur:1981} for more details.

In the supermarket model, when the population size is $n$, we write%
\[
\text{Level }0:\text{ \ }E_{0}^{n}=\left\{  \left(  0,i\right)  :1\leq i\leq
m_{A}\right\}
\]
and for $k\geq1$%
\[
\text{Level }k:\text{ \ }E_{k}^{n}=\left\{  \left(  k,i,j\right)  :1\leq i\leq
m_{A},1\leq j\leq m_{B}\right\}  ,
\]%
\[
E_{n}=\bigcup\limits_{k=0}^{n}\left\{  \text{Level }k\right\}  =\bigcup
\limits_{k=0}^{n}\left\{  E_{k}^{n}\right\}  .
\]
In the state space $E_{n}$, the density dependent jump Markov process for the
supermarket model with MAPs and PH service times contains four classes of
state transitions as follows:

Class one $E_{k}^{n}$ $\underrightarrow{a}$ $E_{k}^{n}$: $\left(  0,i\right)
\rightarrow\left(  0,i^{\ast}\right)  $ or $\left(  k,i,j\right)
\rightarrow\left(  k,i^{\ast},j\right)  $, where $1\leq i,i^{\ast}\leq m_{A}$;

Class two $E_{k}^{n}$ $\underrightarrow{s}$ $E_{k}^{n}$: $\left(
k,i,j\right)  \rightarrow\left(  k,i,j^{\ast}\right)  $, where $1\leq
j,j^{\ast}\leq m_{B}$;

Class three $E_{k}^{n}$ $\underrightarrow{a}$ $E_{k+1}^{n}$: $\left(
0,i\right)  \rightarrow\left(  1,i,j\right)  $ or $\left(  k,i,j\right)
\rightarrow\left(  k+1,i,j\right)  $; and

Class four $E_{k}^{n}$ $\underrightarrow{s}$ $E_{k-1}^{n}$: $\left(
1,i,j\right)  \rightarrow\left(  0,i\right)  $ or $\left(  k,i,j\right)
\rightarrow\left(  k-1,i,j\right)  $.

\noindent Note that the transitions $\underrightarrow{a}$ and
$\underrightarrow{s}$ express arrival transition and service transition, respectively.

We write%
\[
s_{0}^{\left(  i\right)  }\left(  n\right)  =\left(  \frac{0}{n},i\right)  ,
\]%

\[
S_{0}\left(  n\right)  =\left(  s_{0}^{\left(  1\right)  }\left(  n\right)
,s_{0}^{\left(  2\right)  }\left(  n\right)  ,\ldots,s_{0}^{\left(
m_{A}\right)  }\left(  n\right)  \right)  ;
\]
and for $k\geq1$%
\[
s_{k}^{\left(  i,j\right)  }\left(  n\right)  =\left(  \frac{k}{n}%
,i,j\right)
\]
and%
\[
S_{k}\left(  n\right)  =\left(  s_{k}^{\left(  1,1\right)  }\left(  n\right)
,s_{k}^{\left(  1,2\right)  }\left(  n\right)  ,\ldots,s_{k}^{\left(
1,m_{B}\right)  }\left(  n\right)  ;\ldots;s_{k}^{\left(  m_{A},1\right)
}\left(  n\right)  ,s_{k}^{\left(  m_{A},2\right)  }\left(  n\right)
,\ldots,s_{k}^{\left(  m_{A},m_{B}\right)  }\left(  n\right)  \right)  .
\]
Note that the states of the density dependent jump Markov process can be
normalized and interpreted as measuring population densities%
\[
S\left(  n\right)  =\left\{  S_{0}\left(  n\right)  ,S_{1}\left(  n\right)
,S_{2}\left(  n\right)  ,\ldots\right\}  ,
\]
the transition rates of the Markov process depend only on these densities.

Let $\left\{  \widehat{X}_{n}\left(  t\right)  :t\geq0\right\}  $ be a density
dependent jump Markov process on the state space $E_{n}$ whose transition
rates corresponding to the above four cases are given by%
\[
q_{\left(  0,i\right)  \rightarrow\left(  0,i^{\ast}\right)  }^{\left(
n\right)  }=n\beta_{i\rightarrow i^{\ast}}\left(  \frac{0}{n},i\right)
=n\beta_{i\rightarrow i^{\ast}}\left(  s_{0}^{\left(  i\right)  }\left(
n\right)  \right)  ,
\]%
\[
q_{\left(  k,i,j\right)  \rightarrow\left(  k,i^{\ast},j\right)  }^{\left(
n\right)  }=n\beta_{i\rightarrow i^{\ast}}\left(  \frac{k}{n},i,j\right)
=n\beta_{i\rightarrow i^{\ast}}\left(  s_{k}^{\left(  i,j\right)  }\left(
n\right)  \right)  ;
\]%
\[
q_{\left(  k,i,j\right)  \rightarrow\left(  k,i,j^{\ast}\right)  }^{\left(
n\right)  }=n\beta_{j\rightarrow j^{\ast}}\left(  \frac{k}{n},i,j\right)
=n\beta_{j\rightarrow j^{\ast}}\left(  s_{k}^{\left(  i,j\right)  }\left(
n\right)  \right)  ;
\]%
\[
q_{\left(  0,i\right)  \rightarrow\left(  1,i,j\right)  }^{\left(  n\right)
}=n\beta_{0\rightarrow1}\left(  \frac{0}{n},i;j\right)  =n\beta_{0\rightarrow
1}\left(  s_{0}^{\left(  i\right)  }\left(  n\right)  ,j\right)  ,
\]%
\[
q_{\left(  k,i,j\right)  \rightarrow\left(  k+1,i,j\right)  }^{\left(
n\right)  }=n\beta_{k\rightarrow k+1}\left(  \frac{k}{n},i,j\right)
=n\beta_{k\rightarrow k+1}\left(  s_{k}^{\left(  i,j\right)  }\left(
n\right)  \right)  ;
\]%
\[
q_{\left(  1,i,j\right)  \rightarrow\left(  0,i\right)  }^{\left(  n\right)
}=n\beta_{1\rightarrow0}\left(  \frac{1}{n},i,j\right)  =n\beta_{1\rightarrow
0}\left(  s_{1}^{\left(  i,j\right)  }\left(  n\right)  \right)  ,
\]%
\[
q_{\left(  k,i,j\right)  \rightarrow\left(  k-1,i,j\right)  }^{\left(
n\right)  }=n\beta_{k\rightarrow k-1}\left(  \frac{k}{n},i,j\right)
=n\beta_{k\rightarrow k-1}\left(  s_{k}^{\left(  i,j\right)  }\left(
n\right)  \right)  .
\]
Let%
\[
Q_{0}^{\left(  n\right)  }=\left(  \beta_{i\rightarrow i^{\ast}}\left(
s_{0}^{\left(  i\right)  }\left(  n\right)  \right)  \right)  _{1\leq
i,i^{\ast}\leq m_{A}},
\]%
\[
Q_{a,k}^{\left(  n\right)  }=\left(  \beta_{i\rightarrow i^{\ast}}\left(
s_{k}^{\left(  i,j\right)  }\left(  n\right)  \right)  \right)  _{1\leq
i,i^{\ast}\leq m_{A},1\leq j\leq m_{B}};
\]%
\[
Q_{s,k}^{\left(  n\right)  }=\left(  \beta_{j\rightarrow j^{\ast}}\left(
s_{k}^{\left(  i,j\right)  }\left(  n\right)  \right)  \right)  _{1\leq i\leq
m_{A},1\leq j,j^{\ast}\leq m_{B}};
\]%
\[
Q_{0}^{\left(  n\right)  }\left(  A\right)  =\left(  \beta_{0\rightarrow
1}\left(  s_{0}^{\left(  i\right)  }\left(  n\right)  ,j\right)  \right)
_{1\leq i\leq m_{A},1\leq j\leq m_{B}},
\]%
\[
Q_{k}^{\left(  n\right)  }\left(  A\right)  =\left(  \beta_{k\rightarrow
k+1}\left(  s_{k}^{\left(  i,j\right)  }\left(  n\right)  \right)  \right)
_{1\leq i\leq m_{A},1\leq j\leq m_{B}};
\]%
\[
Q_{1}^{\left(  n\right)  }\left(  S\right)  =\left(  \beta_{1\rightarrow
0}\left(  s_{1}^{\left(  i,j\right)  }\left(  n\right)  \right)  \right)
_{1\leq i\leq m_{A},1\leq j\leq m_{B}},
\]%
\[
Q_{k}^{\left(  n\right)  }\left(  S\right)  =\left(  \beta_{k\rightarrow
k-1}\left(  s_{k}^{\left(  i,j\right)  }\left(  n\right)  \right)  \right)
_{1\leq i\leq m_{A},1\leq j\leq m_{B}}.
\]
In the supermarket model, $\widehat{X}_{n}\left(  t\right)  $ is an unscaled
process which records the number of servers with at least $k$ customers for
$0\leq k\leq n$. We write%
\[
Q^{\left(  n\right)  }=\left(
\begin{array}
[c]{ccccc}%
Q_{0}^{\left(  n\right)  } & Q_{0}^{\left(  n\right)  }\left(  A\right)  &  &
& \\
Q_{1}^{\left(  n\right)  }\left(  S\right)  & Q_{a,1}^{\left(  n\right)
}+Q_{s,1}^{\left(  n\right)  } & Q_{1}^{\left(  n\right)  }\left(  A\right)  &
& \\
& Q_{2}^{\left(  n\right)  }\left(  S\right)  & Q_{a,2}^{\left(  n\right)
}+Q_{s,2}^{\left(  n\right)  } & Q_{2}^{\left(  n\right)  }\left(  A\right)  &
\\
&  & \ddots & \ddots & \ddots
\end{array}
\right)  ,
\]
where
\[
Q_{0}^{\left(  n\right)  }=\gamma^{T}\left[  S_{0}\left(  n\right)  \right]
^{\odot d}C,
\]%
\[
Q_{a,k}^{\left(  n\right)  }=\left(  \gamma\otimes\tau\right)  ^{T}\left[
S_{k}\left(  n\right)  \right]  ^{\odot d}\left(  C\otimes I\right)  ;
\]%
\[
Q_{s,k}^{\left(  n\right)  }=\left(  \gamma\otimes\tau\right)  ^{T}%
S_{k}\left(  n\right)  \left(  I\otimes T\right)  ;
\]%
\[
Q_{0}^{\left(  n\right)  }\left(  A\right)  =\gamma^{T}\left[  S_{0}\left(
n\right)  \right]  ^{\odot d}\left(  D\otimes\alpha\right)  ,
\]%
\[
Q_{k}^{\left(  n\right)  }\left(  A\right)  =\left(  \gamma\otimes\tau\right)
^{T}\left[  S_{k}\left(  n\right)  \right]  ^{\odot d}\left(  D\otimes
I\right)  ;
\]%
\[
Q_{1}^{\left(  n\right)  }\left(  S\right)  =\left(  \gamma\otimes\tau\right)
^{T}S_{1}\left(  n\right)  \left(  I\otimes T^{0}\right)  ,
\]%
\[
Q_{k}^{\left(  n\right)  }\left(  S\right)  =\left(  \gamma\otimes\tau\right)
^{T}S_{k}\left(  n\right)  \left(  I\otimes T^{0}\alpha\right)  .
\]

Using Chapter 7 in Kurtz \cite{Kur:1981} or Subsection 3.4.1 in Mitzenmacher
\cite{Mit:1996b}, the Markov process $\left\{  \widehat{X}_{n}\left(
t\right)  :t\geq0\right\}  $ with transition rate matrix $\mathcal{Q}^{\left(
n\right)  }=$ $nQ^{\left(  n\right)  }$ is given by%
\begin{equation}
\widehat{X}_{n}\left(  t\right)  =\widehat{X}_{n}\left(  0\right)  +\sum
_{b\in\mathbb{E}}l_{b}Y_{b}\left(  n\int_{0}^{t}\beta_{l}\left(
\frac{\widehat{X}_{n}\left(  u\right)  }{n}\right)  \text{d}u\right)  ,
\label{Equ6.1}%
\end{equation}
where $Y_{b}\left(  x\right)  $ for $b\in\mathbb{E}$ are independent standard
Poisson processes, $l_{b}$ is a positive integer with $l_{b}\leq\Re<+\infty$,
and%
\begin{align*}
\mathbb{E}=  &  \{\left(  0,i\right)  \rightarrow\left(  0,i^{\ast}\right)
,\left(  k,i,j\right)  \rightarrow\left(  k,i^{\ast},j\right)  ,\left(
k,i,j\right)  \rightarrow\left(  k,i,j^{\ast}\right)  ;\left(  0,i\right)
\rightarrow\left(  1,i,j\right)  ,\\
&  \left(  k,i,j\right)  \rightarrow\left(  k+1,i,j\right)  \text{ for }%
k\geq1;\left(  1,i,j\right)  \rightarrow\left(  0,i\right)  ,\left(
l,i,j\right)  \rightarrow\left(  l-1,i,j\right)  \text{ for }l\geq2\}
\end{align*}
Clearly, the jump Markov process in Equation (\ref{Equ6.1}) at time $t$ is
determined by the starting point and the transition rates which are integrated
over its history.

Let%
\begin{equation}
F\left(  y\right)  =\sum_{b\in\mathbb{E}\left(  y\right)  }l_{b}\beta
_{b}\left(  y\right)  , \label{Equ6.2}%
\end{equation}
where%
\[
\mathbb{E}\left(  y\right)  =\left\{  b\in\mathbb{E}:\text{ the transition
}b\text{ begins from state }y\right\}  .
\]
Taking $X_{n}\left(  t\right)  =n^{-1}\widehat{X}_{n}\left(  t\right)  $ which
is an appropriate scaled process, we have%
\begin{equation}
X_{n}\left(  t\right)  =X_{n}\left(  0\right)  +\sum_{b\in\mathbb{E}}%
l_{b}n^{-1}\widehat{Y}_{b}\left(  n\int_{0}^{t}\beta_{b}\left(  X_{n}\left(
u\right)  \right)  \text{d}u\right)  +\int_{0}^{t}F\left(  X_{n}\left(
u\right)  \right)  \text{d}u, \label{Equ6.2-1}%
\end{equation}
where $\widehat{Y}_{b}\left(  y\right)  =Y_{b}\left(  y\right)  -y$ is a
Poisson process centered at its expectation.

Let $X\left(  t\right)  =\lim_{n\rightarrow\infty}X_{n}\left(  t\right)  $ and
$x_{0}=\lim_{n\rightarrow\infty}X_{n}\left(  0\right)  $, we obtain%
\begin{equation}
X\left(  t\right)  =x_{0}+\int_{0}^{t}F\left(  X\left(  u\right)  \right)
\text{d}u,\text{ \ }t\geq0, \label{Equ6.2-2}%
\end{equation}
due to the fact that%
\[
\lim_{n\rightarrow\infty}\frac{1}{n}\widehat{Y}_{b}\left(  n\int_{0}^{t}%
\beta_{b}\left(  X_{n}\left(  u\right)  \right)  \text{d}u\right)  =0
\]
by means of the law of large numbers. In the supermarket model, the
deterministic and continuous process $\left\{  X\left(  t\right)
,t\geq0\right\}  $ is described by the infinite-size system of differential
vector equations (\ref{Equ3}) to (\ref{Equ6}), or simply,%
\begin{equation}
\frac{d}{dt}X\left(  t\right)  =F\left(  X\left(  t\right)  \right)
\label{Equ6.3}%
\end{equation}
with the initial condition%
\begin{equation}
X\left(  0\right)  =x_{0}. \label{Equ6.4}%
\end{equation}

Now, we consider the uniqueness of the limiting deterministic process
$\left\{  X\left(  t\right)  ,t\geq0\right\}  $ with (\ref{Equ6.3}) and
(\ref{Equ6.4}),\ or the uniqueness of the solution to the infinite-size system
of differential vector equations (\ref{Equ3}) to (\ref{Equ6}). To that end, a
sufficient condition is Lipschitz, that is, for some constant $M>0,$%
\[
|F\left(  y\right)  -F\left(  z\right)  |\leq M||y-z||.
\]
In general, the Lipschitz condition is standard and sufficient for the
uniqueness of the solution to the finite-size system of differential vector
equations; while for the countable infinite-size case, readers may refer to
Theorem 3.2 in Deimling \cite{Dei:1977} and Subsection 3.4.1 in Mitzenmacher
\cite{Mit:1996b} for some useful generalization.

To check the Lipschitz condition, by means of the law of large numbers we have%
\[
\pi_{k}=\lim_{n\rightarrow\infty}S_{k}\left(  n\right)  ,\text{ \ \ }k\geq0,
\]
which leads to%
\begin{equation}
Q=\left(
\begin{array}
[c]{ccccc}%
Q_{0} & Q_{0}\left(  A\right)  &  &  & \\
Q_{1}\left(  S\right)  & Q_{a,1}+Q_{s,1} & Q_{1}\left(  A\right)  &  & \\
& Q_{2}\left(  S\right)  & Q_{a,2}+Q_{s,2} & Q_{2}\left(  A\right)  & \\
&  & \ddots & \ddots & \ddots
\end{array}
\right)  , \label{Equ6.5}%
\end{equation}
where
\[
Q_{0}=\gamma^{T}\pi_{0}^{\odot d}C,
\]%
\[
Q_{a,k}=\left(  \gamma\otimes\tau\right)  ^{T}\pi_{k}^{\odot d}\left(
C\otimes I\right)  ;
\]%
\[
Q_{s,k}=\left(  \gamma\otimes\tau\right)  ^{T}\pi_{k}\left(  I\otimes
T\right)  ;
\]%
\[
Q_{0}\left(  A\right)  =\gamma^{T}\pi_{0}^{\odot d}\left(  D\otimes
\alpha\right)  ,
\]%
\[
Q_{k}\left(  A\right)  =\left(  \gamma\otimes\tau\right)  ^{T}\pi_{k}^{\odot
d}\left(  D\otimes I\right)  ;
\]%
\[
Q_{1}\left(  S\right)  =\left(  \gamma\otimes\tau\right)  ^{T}\pi_{1}\left(
I\otimes T^{0}\right)  ,
\]%
\[
Q_{k}\left(  S\right)  =\left(  \gamma\otimes\tau\right)  ^{T}\pi_{k}\left(
I\otimes T^{0}\alpha\right)  .
\]

Let%
\[
\zeta_{0}=\frac{\pi_{0}^{\odot d}De}{\pi_{0}e}%
\]
and for $k\geq1$%
\[
\zeta_{k}=\frac{\pi_{k}^{\odot d}\left(  De\otimes e\right)  }{\pi_{k}e},
\]%
\[
\eta_{k}=\frac{\pi_{k}\left(  I\otimes T^{0}\alpha\right)  }{\pi_{k}e}.
\]
Then $\zeta_{k},\eta_{k}>0$ for $k\geq1$.

The following theorem shows that the supermarket model with MAPs and PH
service times satisfies the Lipschitz condition for analyzing the uniqueness
of the solution to the infinite-size system of differential vector equations
(\ref{Equ3}) to (\ref{Equ6}).

\begin{The}
\label{The:Lip}The supermarket model with MAPs and PH service times satisfies
the Lipschitz condition.
\end{The}

\textbf{Proof} \ Let the state space of the Markov process $\left\{  X\left(
t\right)  ,t\geq0\right\}  $ be%
\[
\Omega=\left\{  \pi_{k}:k\geq0\right\}  .
\]
For two arbitrary entries $y,z\in\Omega$, we have%
\[
|F\left(  y\right)  -F\left(  z\right)  |\leq\sum_{a\in\mathbb{E}\left(
y\right)  \cap\mathbb{E}\left(  z\right)  }l_{a}|\beta_{a}\left(  y\right)
-\beta_{a}\left(  z\right)  |\leq\Re\sum_{a\in\mathbb{E}\left(  y\right)
\cap\mathbb{E}\left(  z\right)  }|\beta_{a}\left(  y\right)  -\beta_{a}\left(
z\right)  |.
\]
Note that $a$ expresses either an arrival transition or a service transition
in the above four cases. When $a$ expresses an arrival transition, we can
analyze the function $\beta_{a}\left(  y\right)  $ from the two cases of
arrival transitions; while when $b$ expresses a service transition, the
function $\beta_{b}\left(  y\right)  $ can similarly be dealt with from the
two cases of service transitions.

When $a$ expresses an arrival transition, we analyze the function $\beta
_{a}\left(  y\right)  $ based on $a\in\mathbb{E}\left(  y\right)
\cap\mathbb{E}\left(  z\right)  $ from the following two cases.

Case one: $y=\pi_{0},z=\pi_{1}$. In this case, we have%
\begin{align*}
|\beta_{a}\left(  y\right)  -\beta_{a}\left(  z\right)  |  &  =|\pi_{0}^{\odot
d}Ce-\pi_{0}^{\odot d}\left(  D\otimes\alpha\right)  e-\pi_{1}^{\odot
d}\left(  C\otimes I\right)  e|\\
&  =|-\pi_{0}^{\odot d}De-\pi_{0}^{\odot d}De+\pi_{1}^{\odot d}\left(
De\otimes e\right)  |\\
&  =|2\zeta_{0}\pi_{0}e-\zeta_{1}\pi_{1}e|\\
&  =|2\zeta_{0}-\zeta_{1}\pi_{1}e|.
\end{align*}
Taking%
\[
M_{a}\left(  0,1\right)  \geq\frac{|2\zeta_{0}-\zeta_{1}\pi_{1}e|}{1-\pi_{1}%
e},
\]
it is clear that%
\[
|\beta_{a}\left(  y\right)  -\beta_{a}\left(  z\right)  |\leq M_{a}\left(
0,1\right)  \left(  1-\pi_{1}e\right)  =M_{a}\left(  0,1\right)  \left(
\pi_{0}e-\pi_{1}e\right)  .
\]
Note that $\pi_{0}$ and $\pi_{1}$ are two row vectors of sizes $m_{A}$ and
$m_{A}m_{B}$, respectively, in this case we write%
\[
||y-z||=\pi_{0}e-\pi_{1}e.
\]
Thus we have%
\[
|\beta_{a}\left(  y\right)  -\beta_{a}\left(  z\right)  |\leq M_{a}\left(
0,1\right)  ||y-z||.
\]

Case two: $y=\pi_{k-1},z=\pi_{k}$ for $k\geq2$. In this case, we have%
\begin{align*}
|\beta_{a}\left(  y\right)  -\beta_{a}\left(  z\right)  |=  &  |\pi
_{k-2}^{\odot d}\left(  D\otimes I\right)  e+\pi_{k-1}^{\odot d}\left(
C\otimes I\right)  e\\
&  -\pi_{k-1}^{\odot d}\left(  t\right)  \left(  D\otimes I\right)  e-\pi
_{k}^{\odot d}\left(  C\otimes I\right)  e|\\
=  &  |\pi_{k-2}^{\odot d}\left(  t\right)  \left(  De\otimes e\right)
-\pi_{k-1}^{\odot d}\left(  De\otimes e\right) \\
&  -\pi_{k-1}^{\odot d}\left(  De\otimes e\right)  +\pi_{k}^{\odot d}\left(
De\otimes e\right)  |\\
=  &  |\zeta_{k-2}\pi_{k-2}e-2\zeta_{k-1}\pi_{k-1}e+\zeta_{k}\pi_{k}e|.
\end{align*}
Let%
\[
M_{a}\left(  k-1,k\right)  \geq\frac{|\zeta_{k-2}\pi_{k-2}e-2\zeta_{k-1}%
\pi_{k-1}e+\zeta_{k}\pi_{k}e|}{||\pi_{k-1}-\pi_{k}||}.
\]
Then%
\[
|\beta_{a}\left(  y\right)  -\beta_{a}\left(  z\right)  |\leq M_{a}\left(
k-1,k\right)  ||\pi_{k-1}-\pi_{k}||.
\]
Based on the above two cases, taking%
\[
M_{a}=\sup\left\{  M_{a}\left(  0,1\right)  ,M_{a}\left(  k-1,k\right)
:k\geq2\right\}  ,
\]
we obtain that for two arbitrary entries $y,z\in\Omega,$%
\begin{equation}
|\beta_{a}\left(  y\right)  -\beta_{a}\left(  z\right)  |\leq M_{a}||y-z||.
\label{Equ6.6}%
\end{equation}

Similarly, when $b$ expresses a service transition, we can choose a positive
number $M_{b}$ such that for two arbitrary entries $y,z\in\Omega,$%
\begin{equation}
|\beta_{b}\left(  y\right)  -\beta_{b}\left(  z\right)  |\leq M_{b}||y-z||.
\label{Equ6.7}%
\end{equation}

Let $M=\Re\max\left\{  M_{a},M_{b}\right\}  $. Then it follows from
(\ref{Equ6.6}) and (\ref{Equ6.7}) that for two arbitrary entries $y,z\in
\Omega,$%
\[
|F\left(  y\right)  -F\left(  z\right)  |\leq M||y-z||.
\]
This completes the proof. \nopagebreak    \hspace*{\fill} \nopagebreak
\textbf{{\rule{0.2cm}{0.2cm}}}

Based on Theorem \ref{The:Lip}, the following theorem easily follows from
Theorem 3.13 in Mitzenmacher \cite{Mit:1996b}.

\begin{The}
\label{The:Kurtz}In the supermarket model with MAPs and PH service times,
$\left\{  X_{n}\left(  t\right)  \right\}  $ and $\left\{  X\left(  t\right)
\right\}  $ are respectively given by (\ref{Equ6.2-1}) and (\ref{Equ6.2-2}),
we have%
\[
\lim_{n\rightarrow\infty}\sup_{u\leq t}|X_{n}\left(  u\right)  -X\left(
u\right)  |=0,\text{ \ }a.s.
\]
\end{The}

\textbf{Proof} \ It has been shown that in the supermarket model with MAPs and
PH service times, the function $F\left(  y\right)  $ for $y\in\Omega$
satisfies the Lipschitz condition. At the same time, it is easy to take a
subset $\Omega^{\ast}\subset\Omega$ such that%
\[
\left\{  X\left(  u\right)  :u\leq t\right\}  \subset\Omega^{\ast}%
\]
and%
\[
\sup_{\substack{y\in\Omega^{\ast}\\a\in\mathbb{E}\left(  y\right)  }}\beta
_{a}\left(  y\right)  +\sup_{\substack{y\in\Omega^{\ast}\\b\in\mathbb{E}%
\left(  y\right)  }}\beta_{b}\left(  y\right)  <+\infty,
\]
where $a$ and $b$\ express an arrival transition and a service transition,
respectively. Thus, this proof can easily be completed by means of Theorem
3.13 in Mitzenmacher \cite{Mit:1996b}. This completes the proof. \nopagebreak
\hspace*{\fill} \nopagebreak    \textbf{{\rule{0.2cm}{0.2cm}}}

Using Theorem 3.11 in Mitzenmacher \cite{Mit:1996b} and Theorem
\ref{The:Kurtz}, the following theorem for the expected sojourn time that an
arriving tagged customer spends in an initially empty supermarket model with
MAPs and PH service times over the time interval $\left[  0,t\right]  $.

\begin{The}
In the supermarket model with MAPs and PH service times, the expected sojourn
time that an arriving tagged customer spends in an initially empty system over
the time interval $\left[  0,t\right]  $ is bounded above by%
\[
\theta^{d^{2}}\left(  \theta\omega\rho\right)  ^{d}\left(  \tau-\alpha\right)
\left(  -T\right)  ^{-1}e+\frac{1}{\mu}\sum_{k=0}^{\infty}\theta^{d^{k+1}%
}\left(  \theta\omega\rho\right)  ^{\frac{d^{k+1}-d}{d-1}}+o\left(  1\right)
,
\]
where $o\left(  1\right)  $ is understood as $n\rightarrow\infty$.
\end{The}

\section{Concluding remarks}

In this paper, we provide a novel matrix-analytic approach for studying doubly
exponential solutions of the supermarket models with MAPs and PH service
times. We describe the supermarket model as a system of differential vector
equations, and obtain a closed-form solution with a doubly exponential
structure to the fixed point of the system of differential vector equations.
Based on this, we shows that the fixed point can be decomposed into the
product of two factors inflecting arrival information and service information,
and indicate that the doubly exponential solution to the fixed point is not
always unique for more general supermarket models. Furthermore, we analyze the
exponential convergence of the current location of the supermarket model to
its fixed point, and apply the Kurtz Theorem to study density dependent jump
Markov process given in the supermarket model with MAPs and PH service times,
which leads to the Lipschitz condition under which the fraction measure of the
supermarket model weakly converges the system of differential vector
equations. Therefore, we gain a new and crucial understanding of how the
workload probing can help in load balancing jobs with either non-Poisson
arrivals or non-exponential service times.

Our approach given in this paper is useful in the study of load balancing in
data centers and multi-core servers systems. We expect that this approach will
be applicable to the study of other randomized load balancing schemes, for
example, analyzing a renewal arrival process or a general service time
distribution, discussing retrial service discipline and processor-sharing
discipline, and studying supermarket networks.

\section*{Acknowledgements}

The author are very grateful to Professors {\AA}ke Blomqvist and Juan Eloy
Ruiz-Castro whose comments have greatly improved the presentation of this
paper. John C.S. Lui was supported by the RGC grant. The work of Q.L. Li was
supported by the National Science Foundation of China under grant No. 10871114
and the National Grand Fundamental Research 973 Program of China under grant
No. 2006CB805901.


\vskip                0.2cm

\begin{thebibliography}{9}                                                                                                %

\bibitem {Adl:1998}R. Adler, R. Feldman and M.S. Taqqu (1998). \textit{A
Practical Guide to Heavy Tails: Statistical Techniques for Analyzing Heavy
Tailed Distributions}. Birkh\"{a}user: Boston.

\bibitem {And:1998}A.T. Andersen and B.F. Nielsen (1998). A Markovian approach
for modeling packet traffic with long-range dependence. \textit{IEEE Journal
on Selected Areas in Communications} \textbf{16}, 719--732.

\bibitem {Azar:1999}Y. Azar, A.Z. Broder, A.R. Karlin and E. Upfal (1999).
Balanced allocations. \textit{SIAM Journal on Computing} \textbf{29}, 180--200.

\bibitem {Bra:2010}M. Bramson, Y. Lu and B. Prabhakar (2010). Randomized load
balancing with general service time distributions. In \textit{Proceedings of
the ACM SIGMETRICS international conference on Measurement and modeling of
computer systems}, pages 275--286.

\bibitem {Cha:2000}S.R. Chakravarthy (2000). The Batch Markovian Arrival
Process: A Review and Future Work. In \textit{Advances in Probability Theory
and Stochastic Processes}, A. Krishnamoorthy, N. Raju and V. Ramaswami (eds),
Notable Publications: New Jersey, pages 21--39.

\bibitem {Cor:2009}J.D. Cordeiro and J.P. Kharoufeh (2009). Batch Markovian
Arrival Processes (BMAP). Research Report.

\bibitem {Dah:1999}M. Dahlin (1999). Interpreting stale load information.
\textit{IEEE Transactions on Parallel and Distributed Systems} \textbf{11}, 1033--1047.

\bibitem {Dei:1977}K. Deimling (1977). \textit{Ordinary Differential Equations
in Banach Spaces}. Springer-Verlag.

\bibitem {Eag:1986a}D.L. Eager, E.D. Lazokwska and J. Zahorjan (1986).
Adaptive load sharing in homogeneous distributed systems. \textit{IEEE
Transactions on Software Engineering} \textbf{12}, 662--675.

\bibitem {Eag:1986b}D.L. Eager, E.D. Lazokwska and J. Zahorjan (1986). A
comparison of receiver-initiated and sender-initiated adaptive load sharing.
\textit{Performance Evaluation Review} \textbf{6}, 53--68.

\bibitem {Eag:1988}D.L. Eager, E.D. Lazokwska and J. Zahorjan (1988). The
limited performance benefits of migrating active processes for load sharing.
\textit{Performance Evaluation Review} \textbf{16}, 63--72.

\bibitem {Gra:2000a}C. Graham (2000). Kinetic limits for large communication
networks. In \textit{Modelling in Applied Sci-ences}, N. Bellomo and M.
Pulvirenti (eds.), Birkh\"{a}user: Boston, pages. 317--370.

\bibitem {Gra:2000b}Graham, C. (2000). Chaoticity on path space for a queueing
network with selection of the shortest queue among several. \textit{Journal of
Applied Probabability} \textbf{37}, 198--201.

\bibitem {Gra:2004}Graham, C. (2004). Functional central limit theorems for a
large network in which customers join the shortest of several queues.
\textit{Probability Theory Related Fields} \textbf{131}, 97--120.

\bibitem {harchol97}M. Harchol-Balter and A.B. Downey (1997). Exploiting
process lifetime distributions for dynamic load balancing. \textit{ACM
Transactions on Computer Systems} \textbf{15}, 253--285.

\bibitem {Kur:1981}T.G. Kurtz (1981). \textit{Approximation of Population
Processes}. SIAM.

\bibitem {Li:2010}Q.L. Li (2010). \textit{Constructive Computation in
Stochastic Models with Applications: The RG-Factorizations}. Springer and
Tsinghua Press.

\bibitem {Li:2010b}Q.L. Li (2010). Doubly exponential solution for randomized
load balancing models with general service times. Submited for publication.

\bibitem {Li:2010a}Q.L. Li, John C.S. Lui and Y. Wang (2010). A
matrix-analytic solution for randomized load balancing models with phase-type
service times. In \textit{International Workshop on Performance Evaluation of
Computer and Communication Systems}, Lecture Notes of Computer Science, W.
Gansterer, H. Hlavacs and K.A. Hummel (eds), Springer.

\bibitem {Luc:1991}D.M. Lucantoni (1991). New results on the single server
queue with a batch Markovian arrival process. \textit{Stochastic Models}
\textbf{7}, 1--46.

\bibitem {Luc:2006}M. Luczak and C. McDiarmid (2006). On the maximum queue
length in the supermarket model. \textit{The Annals of Probability}
\textbf{34}, 493--527.

\bibitem {Luc:2007}M. Luczak and C. McDiarmid (2007). Asymptotic distributions
and chaos for the supermarket model. \textit{Electronic Journal of
Probability} \textbf{12}, 75--99.

\bibitem {LucN:2005}M.J. Luczak and J.R. Norris (2005). Strong approximation
for the supermarket model. \textit{The Annals of Applied Probability}
\textbf{15}, 2038--2061.

\bibitem {Mar:2001}J.B. Martin (2001). Point processes in fast Jackson
networks. \textit{Annals of Applied Probability} \textbf{11}, 650--663.

\bibitem {Mar:1999}J.B. Martin and Y.M Suhov (1999). Fast Jackson networks.
\textit{Annals of Applied Probability} \textbf{9}, 854--870.

\bibitem {Mit:1996a}M.D. Mitzenmacher (1996). Load balancing and density
dependent jump Markov processes. In \textit{Proceedings of the Thirty-Seventh
Annual Symposium on Foundations of Computer Science}, pages 213--222.

\bibitem {Mit:1996b}M.D. Mitzenmacher (1996). \textit{The power of two choices
in randomized load balancing}. PhD thesis, University of California at
Berkeley, Department of Computer Science, Berkeley, CA, 1996.

\bibitem {Mit:1998a}M. Mitzenmacher (1998). Analyses of load stealing models
using differential equations. In \textit{Proceedings of the Tenth ACM
Symposium on Parallel Algorithms and Architectures}, pages 212--221.

\bibitem {Mit:1999a}M. Mitzenmacher (1999). On the analysis of randomized load
balancing schemes. \textit{Theory of Computing Systems} \textbf{32}, 361--386.

\bibitem {Mit:1999b}M. Mitzenmacher (1999). Studying balanced allocations with
differential equations. \textit{Combinatorics, Probability, and Computing}
\textbf{8}, 473--482.

\bibitem {Mit:2000}M. Mitzenmacher (2000). How useful is old information?
\textit{IEEE Transactions on Parallel and Distributed Systems} \textbf{11}, 6--20.

\bibitem {Mit:2001}M. Mitzenmacher (2001). The power of two choices in
randomized load balancing. \textit{IEEE Transactions on Parallel and
Distributed Computing} \textbf{12}, 1094--1104.

\bibitem {Mit:2001a}M. Mitzenmacher, A. Richa, and R. Sitaraman (2001). The
power of two random choices: a survey of techniques and results. In
\textit{Handbook of Randomized Computing: volume 1}, P. Pardalos, S.
Rajasekaran and J. Rolim (eds), pages 255--312.

\bibitem {Mit:1998}M. Mitzenmacher and B. V\"{o}cking (1998). The asymptotics
of selecting the shortest of two, improved. In \textit{Proceedings of the 37th
Annual Allerton Conference on Communication, Control, and Computing}, pages 326--327.

\bibitem {Mir:1989}R. Mirchandaney, D. Towsley, and J.A. Stankovic (1989).
Analysis of the effects of delays on load sharing. \textit{IEEE Transactions
on Computers} \textbf{38}, 1513--1525.

\bibitem {Neu:1981}M.F. Neuts (1981). \textit{Matrix-Geometric Solutions in
Stochastic Models-An Algorithmic Approach,} The Johns Hopkins University
Press: Baltimore.

\bibitem {Neu:1989}M.F. Neuts (1989). \textit{Structured stochastic matrices
of }$M/G/1$\textit{ type and their applications.} Marcel Decker Inc.: New York.

\bibitem {Neu:1993}M.F. Neuts (1993). The burstiness of point processes.
\textit{Stochastic Models} \textbf{9}, 445--466

\bibitem {Neu:1995}M.F. Neuts (1995). Matrix-analytic methods in the theory of
queues. In \textit{Advances in queueing: Theory, methods and open problems},
J.H. Dshalalow (ed), 265--292.

\bibitem {Suh:2002}Y.M. Suhov and N.D. Vvedenskaya (2002). Fast Jackson
Networks with Dynamic Routing. \textit{Problems of Information Transmission}
\textbf{38}, 136--153.

\bibitem {Voc:1999}B. V\"{o}cking (1999). How asymmetry helps load balancing.
In \textit{Proceedings of the Fortieth Annual Symposium on Foundations of
Computer Science}, pages 131--140.

\bibitem {Vve:1996}N.D. Vvedenskaya, R.L. Dobrushin and F.I. Karpelevich
(1996). Queueing system with selection of the shortest of two queues: An
asymptotic approach. \textit{Problems of Information Transmissions}
\textbf{32}, 20--34.

\bibitem {Vve:1997}N.D. Vvedenskaya and Y.M. Suhov (1997). Dobrushin's
mean-field approximation for a queue with dynamic routing. \textit{Markov
Processes and Related Fields} \textbf{3}, 493--526.

\bibitem {Web:1978}R. Weber (1978). On the optimal assignment of customers to
parallel servers. \textit{Journal of Applied Probabiblities} \textbf{15}, 406--413.

\bibitem {Win:1977}W. Winston (1977). Optimality of the shortest line
discipline. \textit{Journal of Applied Probabilities} \textbf{14}, 181--189.

\bibitem {Yos:2001}T. Yoshihara, S. Kasahara and Y. Takahashi (2001).
Practical time-scale fitting of self-similar traffic with Markov-modulated
Poisson process. \textit{Telecommunication Systems} \textbf{3}, 185--211.

\bibitem {Zhou:1988}S. Zhou (1988). A trace-driven simulation study of dynamic
load balancing. \textit{IEEE Transactions on Software Engineering }%
\textbf{14}, 1327--1341.
\end{thebibliography}
\end{document}